\documentclass[twocolumn,showpacs,preprintnumbers,amsmath,amssymb,prb]{revtex4-1}

\usepackage{graphicx}
\usepackage{color}

\begin{document}

\title{Nuclear quantum effects in graphene bilayers}
\author{Carlos P. Herrero}
\author{Rafael Ram\'irez}
\affiliation{Instituto de Ciencia de Materiales de Madrid,
         Consejo Superior de Investigaciones Cient\'ificas (CSIC),
         Campus de Cantoblanco, 28049 Madrid, Spain }
\date{\today}

\begin{abstract}
Graphene bilayers display peculiar electronic and mechanical 
characteristics associated to their two-dimensional character 
and relative disposition of the sheets.
Here we study nuclear quantum effects in graphene bilayers
by using path-integral molecular dynamics simulations, which
allow us to consider quantization of vibrational modes
and study the effect of anharmonicity on physical variables.
Finite-temperature properties are analyzed in the range from 
12 to 2000~K.  Our results for graphene bilayers are compared 
with those found for graphene monolayers and graphite.
Nuclear quantum effects turn out to be appreciable in the layer area
and interlayer distance at finite temperatures.
Differences in the behavior of in-plane and real areas
of the graphene sheets are discussed.
The interlayer spacing has a zero-point expansion of 
$1.5 \times 10^{-2}$ \AA\ with respect to the classical minimum.
The compressibility of graphene bilayers in the out-of-plane
direction is found to be similar to that of graphite at low
temperature, and increases faster as temperature is raised.
The low-temperature compressibility increases by a 6\% due to
zero-point motion.
Especial emphasis is laid upon atomic vibrations in the out-of-plane 
direction.  Quantum effects are present in these vibrational modes,
but classical thermal motion becomes dominant over quantum 
delocalization for large system size.
The significance of anharmonicities in this atomic motion is estimated
by comparing with a harmonic approximation for the vibrational
modes in graphene bilayers.
\end{abstract}



\maketitle

\section{Introduction}

In recent years there has been amid scientists a surge of 
interest in two-dimensional (2D) materials, which display
thicknesses in the atomic scale, reaching in some cases 
the level of one atomic layer.
Among these materials, graphene is one of the most studied
due to its remarkable electronic,\cite{ge07,ca09b}
mechanical,\cite{le08,ga18} and 
thermal properties.\cite{gh08b,se10,ba11}
An additional advantage of graphene, as compared with other
layered systems, is that monolayers, bilayers and multilayers 
can be produced in a controlled manner,\cite{zh16b}
which allows one to carry out detailed studies of their
physical properties.

Bilayer graphene, in particular, presents peculiar electronic 
properties different from those of monolayer graphene and 
graphite.\cite{no06}
It is known that it displays ripples and 
out-of-plane deformations like the monolayers,\cite{me07b} 
which are thought to be a relevant scattering mechanism
for electrons in this material.\cite{gi10}  
Recently, graphene bilayers have attracted increased interest
after the discovery that they display unconventional superconductivity
by stacking both sheets twisted relative to each other 
by a small angle.\cite{ca18}
Moreover, it has been lately found the appearance of  
Mott-like insulator states, due to the presence of electrons 
localized in the superlattice associated to a moir\'e
pattern,\cite{ca18b} as well as uncommon magnetic
properties.\cite{yn19}

A deep comprehension of structural and thermodynamic properties of 
2D systems in three-dimensional space has been for many years a
persistent goal in statistical physics.\cite{sa94,ne04}
This question has been largely discussed in the context of biological
membranes and soft condensed matter.\cite{ch15,ru12}
Nevertheless, the complexity of these systems makes it difficult to 
work out microscopic approaches based on realistic interatomic interactions.
Graphene bilayers constitute a singular realization of crystalline 
membranes composed of two atomic layers, where an atomistic description 
is feasible, thus paving the way for a deeper insight into
the physical properties of this type 
of systems.\cite{fa07,ba11,al13,am14,ma16}
Additionally, graphene can give information on the thermodynamic 
stability of 2D crystals in general, which has been debated and may be
related to the anharmonic coupling between in-plane and 
out-of-plane vibrational modes.\cite{am14,co16}

Atomistic simulations have been used to understand finite-temperature
properties of graphene.\cite{fa07,ca09c,ak12,ch14,ma14,lo16}
In general, carbon atoms were treated in
those simulations as classical particles.
However, the Debye temperature of graphene for out-of-plane vibrational 
modes is $\Theta_D^{\rm out} \gtrsim$ 1000~K,\cite{po11}
which suggests that quantum fluctuations should be relevant 
up to relatively high temperatures.
The quantum character of atomic motion can be taken into account
by using path-integral simulations, which allow one to consider
thermal and quantum fluctuations at finite temperatures.\cite{gi88,ce95}
In these methods the nuclear degrees of freedom can be quantized in 
an efficient way, thus permitting to carry out quantitative analyses 
of anharmonic effects in condensed matter.
This kind of simulations were carried out for graphene monolayers 
in recent years to study structural and thermodynamic properties 
of this material.\cite{br15,he16,he18,ha18}

Here we extend those studies to graphene bilayers, for which new
features appear as a consequence of the interaction between layers
and the resultant coupling between atomic displacements
in the out-of-plane direction.
We use the path-integral molecular dynamics (PIMD) method to 
investigate the properties of graphene bilayers at temperatures 
between 12 and 2000~K.
Simulation cells of different sizes are considered, as finite-size 
effects are known to be important for some
properties of graphene.\cite{ga14,he16,lo16}
The thermal behavior of the graphene surface is studied, considering
the difference between in-plane and real areas.
Moreover, the interlayer distance and its thermal fluctuations
yield information about the compressibility of the bilayer 
in the out-of-plane direction.
Our results for the bilayer are compared with those obtained for
graphene monolayers and graphite, which provides information
on the transition of physical properties from an isolated graphene
sheet to the bulk material.
Results of the simulations are also compared with
predictions based on harmonic vibrations of the crystalline sheets.
This approximation happens to be noticeably accurate at low temperatures,
once the frequencies of out-of-plane modes in graphene bilayers 
are properly described.

Path-integral methods similar to that used here have been previously
applied to investigate nuclear quantum effects in pure and doped 
carbon-based materials, such as diamond\cite{ra06} 
and graphite.\cite{he10}   More recently, 
these computational techniques have been also employed to study 
adsorption of helium and hydrogen on graphene.\cite{kw12,he09a,da14} 

 The paper is organized as follows. In Sec.~II, we describe the
computational methods and details of the calculations.
In Sec.~III we present our results for the in-plane and real
area of bilayer graphene. In Sec.~IV we present data of the internal
energy and the different contributions to it.  The interlayer
spacing and compressibility of the bilayer in the out-of-plane
direction is discussed in Sec.~V, and the character of the out-of-plane
atomic motion (classical vs quantum) is dealt with in Sec.~VI.
Finally we summarize the main results in Sec.~VII.

\section{Computational Method}

\subsection{Path-integral molecular dynamics}

We use PIMD simulations to study equilibrium properties of
graphene bilayers at various temperatures.
This procedure is based on the path-integral formulation of statistical
mechanics, which is a suitable nonperturbative approach to study 
finite-temperature properties of many-body quantum systems.\cite{fe72}
In actual applications of this method to numerical simulations,
each quantum particle is described as a set of $N_{\rm Tr}$ beads
(the so-called {\em Trotter number}), which behave as classical-like
particles arranged to build a ring polymer.\cite{gi88,ce95}
This representation becomes formally exact in the limit 
$N_{\rm Tr} \to \infty$.
Details on this kind of simulation techniques can be found
in Refs.~\onlinecite{gi88,ce95,he14,ca17}.

Here we use molecular dynamics simulations to sample the configuration
space of the classical isomorph of our quantum system ($2 N$ carbon atoms).
The dynamics in this computational technique is artificial, as it
does not represent the real quantum dynamics of the actual particles
under consideration. However, it turns out to be very effective to
sample the many-body configuration space, thus yielding accurate results for 
time-independent equilibrium properties of the quantum system.

In the context of PIMD simulations, one needs a Born-Oppenheimer
surface for the nuclear dynamics, which should be derived from 
an adequate description of the interatomic interactions.
Employing an {\em ab-initio} method would largely restrict
the size of the manageable simulation cells, so we use the
LCBOPII effective potential, a long-range bond order potential, mainly
employed to carry out classical simulations of carbon-based
systems.\cite{lo05}
It has been used, in particular, to study the phase diagram of carbon
(diamond, graphite, and liquid carbon), displaying its reliability 
by predicting rather accurately the diamond-graphite
transition line.\cite{gh05b}
In the last few years, the LCBOPII potential was also employed
to reliably describe various properties of graphene,\cite{fa07,lo16}
in particular its Young's modulus.\cite{ra17,za09,po12,ra18b}
This potential has been lately used to perform PIMD simulations, 
aiming at assessing the magnitude of nuclear quantum effects 
in graphene monolayers,\cite{he16} and to study thermodynamic 
properties of this material.\cite{he18}
In this paper, according to previous simulations,\cite{ra16,he16,ra17}
the earliest parameterization of the LCBOPII potential
has been slightly modified to increase the zero-temperature
bending constant $\kappa$ of a graphene monolayer from 
0.82 to 1.49 eV, a value closer to experimental data and
{\em ab-initio} calculations.\cite{la14}

Calculations have been carried out in the isothermal-isobaric ensemble,
where we fix the number of C atoms ($2 N$), the in-plane applied stress 
(here $P_{xy} = 0$), and the temperature ($T$).
We have employed effective algorithms for performing PIMD simulations in 
this ensemble, such as those described in the literature for this kind 
of simulations.\cite{tu98,ma99,tu02}
Staging variables were used to define the bead coordinates, and
the constant-temperature ensemble was accomplished by coupling chains
of four Nos\'e-Hoover thermostats to each staging variable.
An additional chain of four barostats was coupled to the in-plane area 
of the simulation box ($xy$ plane) to give the constant 
pressure $P_{xy} = 0$.\cite{tu98,he14}
The kinetic energy $E_k$ has been calculated by using the {\em virial}
estimator, which displays a statistical uncertainty
much smaller than the {\em primitive} estimator, especially
at high temperatures.\cite{he82,tu98}
This means that the error bar associated to the kinetic energy in
our calculations at $T > 100$~K is smaller than that corresponding to 
the potential energy of the system. The relative accuracy of the kinetic
energy, as compared with the potential energy, increases for
rising temperature. Both error bars are similar at $T =$ 50~K,
whereas at 1000~K and 2000~K the error bar of the kinetic energy is
8 and 25 times less than that of the potential energy, respectively. 
Other technical details about this type of simulations can be found
elsewhere.\cite{he06,he11,ra12}

We have carried out PIMD simulations of graphene bilayers with AB
stacking employing
simulations boxes with $2 N$ carbon atoms, $N$ going from 24 to 8400.
The considered cells had similar side lengths in the $x$ and $y$ directions
($L_x \approx L_y$), for which periodic boundary conditions were assumed.
Carbon atoms can unrestrictedly move in the out-of-plane direction,
thus simulating a free-standing graphene bilayer, i.e.,
we have free boundary conditions in the $z$ coordinate. 
The temperature $T$ was in the range from 12.5 to 2000 K.
For a given temperature, a typical simulation run included
$2 \times 10^5$ PIMD steps for system equilibration, followed by
$8 \times 10^6$ steps for the calculation of average properties.
The Trotter number $N_{\rm Tr}$ was taken proportional to the inverse 
temperature, as $N_{\rm Tr} T$ = 6000~K, which roughly keeps
a constant precision for the PIMD results at different temperatures.
The time step $\Delta t$ for the molecular dynamics has been
taken as 0.5 fs, which is adequate for the C atomic mass and 
the temperatures considered here.
To compare with the results of these simulations, some classical
molecular dynamics simulations of graphene bilayers were also 
performed. In our context this corresponds to setting $N_{\rm Tr}$ = 1.

To compare with our results for graphene bilayers, we have also 
carried out some PIMD simulations of graphite with the same interatomic
potential LCBOPII. In this case we employed cells consisting of
$4 N$ carbon atoms, i.e., four graphene sheets, and periodic
boundary conditions were applied in the three space directions.
We used cells with $N$ = 240 and 960.

\subsection{Layer area}

In the isothermal-isobaric ensemble employed here one fixes 
the applied stress $P_{xy}$ in the $xy$ plane, as indicated above,
which allows for fluctuations in the in-plane area of the simulation
cell, given by $L_x L_y$.
Carbon atoms can unrestrictedly move in the $z$ direction (out-of-plane
direction), so a measure of the {\em real} surface of a graphene sheet 
at finite temperatures will give a value larger than the in-plane area.

The concept of a real surface has been discussed for biological membranes
as an interesting tool to describe some of their physical properties,
instead of the {\em projected} in-plane surface.\cite{wa09,ch15}
Something similar has been proposed in recent years for crystalline 
membranes such as graphene.\cite{po11b,he16,ni17,ra17}
For biological membranes, it was shown that values of
the compressibility may appreciably differ when they are related
to the real area $A$ or in-plane area $A_p$, and something analogous 
has been noticed recently for the elastic properties of graphene 
monolayers, as derived from 
classical molecular dynamics simulations.\cite{ra17}
The real area has been also called true, actual, or
effective area in the literature.\cite{fo08,wa09,ch15}
A clear distinction between both areas is basic to understand some 
thermodynamic properties of 2D materials. 
In fact, the area $A_p$ is the variable conjugate to the in-plane stress 
$P_{xy}$ in the isothermal-isobaric ensemble employed here, whereas the real 
area $A$ is conjugate to the usually-called surface tension.\cite{sa94}
Nicholl {\em et al.}\cite{ni15,ni17} have shown that certain 
experimental techniques are sensitive to properties related to the 
real area $A$, whereas other methods may be suitable to quantify variables
associated to the in-plane area $A_p$.
The difference between both areas, $A - A_p$, has been named 
{\em hidden} area for graphene in Ref.~\onlinecite{ni17}, and 
{\em excess} area in the context of biological membranes.\cite{he84,fo08}

For each graphene sheet,
we calculate the {\em real} area $A$ by a triangulation based on 
the positions of the C atoms along a simulation run. 
In this procedure, $A$ is obtained as a sum of areas of
the hexagons in the graphene structure.\cite{ra17,he18b}
Each hexagon contributes as a sum of six triangles, each one formed
by the positions of two adjacent C atoms and the barycenter of
the hexagon.\cite{ra17,he18b}
In our path-integral method, the area $A$ is given as an average over 
the $N_{\rm Tr}$ beads associated to the atomic nuclei:
\begin{equation}
  A =  \left \langle \frac{1}{N_{\rm Tr}} \sum_{j=1}^{N_{\rm Tr}} A^j
        \right \rangle  \, ,
\label{aa}
\end{equation}
where $A^j$ is the instantaneous area per atom for {\em imaginary time}
(bead) $j$, and the brackets indicate an ensemble average 
(a mean value for a simulation run).

In general, $A \geq A_p$, and both areas coincide for strictly planar
graphene layers, as occurs in a classical calculation at $T = 0$.
When one takes into account nuclear quantum effects, $A$ and $A_p$ 
are not exactly equal, even for $T \to 0$, because of the 
zero-point motion of C atoms in the transverse $z$ direction.
For graphene monolayers it was found that both areas display 
temperature dependencies qualitatively different: 
while $A_p$ shows a negative thermal expansion in a wide temperature 
range, $A$ does not present such a behavior.\cite{za09,he16}
This may be different for graphene bilayers, as discussed below.
In the following, $A$ and $A_p = L_x L_y / N$ will refer to the real 
and in-plane area per atom, respectively.

\subsection{Mean-square displacements}

For each quantum path of a particle (here atomic nucleus),
we define the {\em centroid} (center of mass) as
\begin{equation}
   \overline{\bf r}_i = \frac{1}{N_{\rm Tr}} 
          \sum_{j=1}^{N_{\rm Tr}} {\bf r}_{ij} \; ,
\label{centr}
\end{equation}
where ${\bf r}_{ij} \equiv (x_{ij}, y_{ij}, z_{ij})$ is the 
three-dimensional position
of bead $j$ in the ring polymer associated to nucleus $i$.
For the out-of-plane motion we consider the $z$-coordinate of
the polymer beads.
Then, the mean-square displacement $(\Delta z)^2_i$ of atomic nucleus
$i$ ($i = 1, ..., 2N$) in the $z$ direction along a PIMD simulation run 
is defined as
\begin{equation}
 (\Delta z)^2_i =  \frac{1}{N_{\rm Tr}}  \sum_{j=1}^{N_{\rm Tr}}
     \left<  ( z_{ij} - \left< \overline{z}_i \right>)^2 \right>   \, ,
\label{deltaz2b}
\end{equation}
Here $\overline{z}_i$ is the instantaneous $z$-coordinate of the
centroid of atom $i$, and $\left< \overline{z}_i \right>$ is its
mean value along a simulation run. Hence,
$\left<  ( z_{ij} - \left< \overline{z}_i \right>)^2 \right>$ is
the MSD of the z-coordinate $z_{ij}$ of bead $j$ with respect to the
average centroid $\left< \overline{z}_i \right>$.
Then, $(\Delta z)^2_i$ is the mean of those displacements for the
beads corresponding to atomic nucleus $i$ (j = 1, ..., $N_{\rm Tr}$).

The kinetic energy of a particle is related to its quantum delocalization,
or in the present context, to the spread of the paths associated to it.
This can be measured by the mean-square {\em radius-of-gyration} of
the ring polymers, with an out-of-plane component:\cite{gi88,gi90}
\begin{equation}
  Q_{z,i}^2 = \frac{1}{N_{\rm Tr}}   \sum_{j=1}^{N_{\rm Tr}}
           \left<  (z_{ij} - \overline{z}_i)^2 \right>    \, .
\label{qzi2}
\end{equation}
Note the difference between the r.h.s. of Eqs.~(\ref{deltaz2b}) 
and (\ref{qzi2}): in the former there appears an average of the centroid
position over the whole trajectory, $\left< \overline{z}_i \right>$,
whereas in the latter we have the instantaneous $\overline{z}_i$
for each configuration.
Then, $Q_{z,i}^2$ is the $z$-component of the mean-square 
"radius-of-gyration" of the paths corresponding to atom $i$. 
The total spatial delocalization $(\Delta z)^2_i$ of atomic nucleus $i$ in
the $z$ direction at a finite temperature includes, in addition to
$Q_{z,i}^2$, another contribution which accounts for classical-like
motion of the centroid coordinate $\overline{z}_i$, i.e.
\begin{equation}
    (\Delta z)^2_i = C_{z,i}^2 + Q_{z,i}^2  \, ,
\label{deltaz2c}
\end{equation}
with
\begin{equation}
 C_{z,i}^2 = \left< \left( \overline{z}_i - \langle \overline{z}_i \rangle
                \right)^2 \right>
       =  \langle  \overline{z}_i^2 \rangle -
          \langle  \overline{z}_i \rangle^2  \, .
\label{czi2}
\end{equation}

Thus, in our context of the path-integral formulation, the term 
$C_{z,i}^2º$ is the mean-square displacement (MSD) of the  centroid
of atomic nucleus $i$, and the quantum
component $Q_{z,i}^2$ is the average MSD of the path (here beads)
with respect to the instantaneous centroid. 
$C_{z,i}^2$ behaves as a semiclassical thermal contribution
to $(\Delta z)^2_i$, since at high temperature it converges to
the mean-square displacement corresponding
to a classical model, and in this limit the
quantum paths collapse onto single points ($Q_{z,i}^2 \to 0$).

In the other limit, for $T \to 0$, $C_{z,i}^2$ vanishes and
$Q_{z,i}^2$ corresponds to zero-point
motion of atomic nucleus $i$.
In the results presented below, we will show data for $(\Delta z)^2$
calculated as an average for $2 N$ atoms in the simulation cell:
\begin{equation}
    (\Delta z)^2 = \frac{1}{2N}  \sum_{i=1}^{2N}  (\Delta z)^2_i   \, ,
\end{equation}
and similarly for $Q_z^2$ and $C_z^2$.

\subsection{Harmonic approximation}

For out-of-plane vibrations in graphene bilayers,
the behavior of $C_z^2$ and $Q_z^2$ may be explained in terms of
a harmonic model for the vibrational modes.
In a quantum harmonic approximation (HA), the mean-square displacement 
at temperature $T$ is given by
\begin{equation}
  (\Delta z)^2  =  \frac{1}{2N} \sum_{i,\bf k} 
     \frac{\hbar}{2 m \omega_i({\bf k})}
     \coth \left( \frac{\hbar \omega_i({\bf k})}{2 k_B T}  \right)  \, ,
\label{qz2b}
\end{equation}
where $m$ is the carbon atomic mass, $k_B$ is Boltzmann's constant,
and the index $i$ ($i$ = 1, ..., 4) refers to the phonon bands 
with atomic displacements along the $z$ direction 
(ZA, ZO', and two ZO bands).\cite{ka11,ya08,si13b,ko15b}
The sum in ${\bf k}$ is extended to wavevectors
${\bf k} = (k_x, k_y)$ in the 2D hexagonal Brillouin zone,
with ${\bf k}$ points spaced by $\Delta k_x = 2 \pi / L_x$ and
$\Delta k_y = 2 \pi / L_y$.\cite{ra16}
For rising system size $N$, there appear vibrational modes with
longer wavelength $\lambda$. We have an effective wavelength
cut-off $\lambda_{max} \approx L$, with $L = (N A_p)^{1/2}$,
so that the minimum wavevector accessible is
$k_0 = 2 \pi / \lambda_{max}$, which scales as $k_0 \sim N^{-1/2}$.

The classical contribution to the MSD at temperature $T$ is given by
\begin{equation}
    C_z^2  = \frac{1}{2N} \sum_{i,\bf k}
            \frac{k_B T}{m \omega_i({\bf k})^2}    \, .
\label{cz2b}
\end{equation}
This expression is accurate for classical motion at relatively low 
temperatures, and at high $T$ anharmonicity causes $C_z^2$ to increase 
sublinearly with $T$, as shown below.
From $(\Delta z)^2$ and $C_z^2$, we calculate $Q_z^2$ as the
difference $Q_z^2 = (\Delta z)^2 - C_z^2$.
For $T \to 0$ the MSD converges in the harmonic approximation to
\begin{equation}
	(\Delta z)^2_0 =  Q_{z,0}^2  =
   \frac{1}{2N} \sum_{i, \bf k} \frac{\hbar}{2 m \omega_i({\bf k})} \, .
\label{qz20}
\end{equation}

For the calculations presented below, based on the HA,
we have used the vibrational frequencies of the four phonon   
branches in the $z$ direction, derived from a diagonalization of 
the dynamical matrix corresponding to the LCBOPII potential 
employed here.
The main difference with the phonon bands in monolayer graphene is 
the appearance of the layer-breathing ZO' band, which is nearly 
flat in the 2D ${\bf k}$-space region
close to the $\Gamma$ point ($|{\bf k}| = k = 0$),
with a frequency for small $k$: $\omega_0$ = 92 cm$^{-1}$.
This value is close to that obtained earlier from {\em ab-initio}
calculations for graphene bilayers.\cite{ya08}

\section{In-plane and real area}

For an isolated graphene layer we find in a classical calculation
in the $T = 0$ limit a planar sheet with an interatomic
distance $d_{\rm C-C}$ = 1.4199 \AA. 
In a quantum approach, the low-temperature limit includes out-of-plane
zero-point motion, so that the graphene
sheet is not strictly plane even at $T = 0$.
In addition to this, anharmonicity of in-plane vibrations
gives rise to a zero-point bond expansion,
yielding an interatomic distance of 1.4287 \AA, i.e., an increase of
$8.8 \times 10^{-3}$~\AA\ with respect to the classical value.\cite{he16}

For the graphene bilayer we find at the minimum-energy configuration
$d_{\rm C-C}$ = 1.4193~\AA, a little smaller than for the monolayer,
similarly to the results presented in Ref.~\onlinecite{za10b}.
Our PIMD simulations give for the bilayer a low-temperature
C--C distance of 1.4281~\AA, i.e., the same zero-point expansion of
the interatomic distance as that found for the graphene monolayer.
Even if this increase in bond length may seem small, it is much larger
than the precision reached for determining cell parameters from
diffraction techniques.\cite{ya94,ra93b,ka98}
The bond expansion due to nuclear quantum effects decreases as
temperature is raised, as in the high-$T$ limit the
classical and quantum predictions have to converge one to the other.
However, at temperatures so high as 2000~K the value of $d_{\rm C-C}$
derived from PIMD simulations is still distinctly larger than
the classical prediction.

As indicated in Sec.~II, in the isothermal-isobaric ensemble employed 
here we allow for fluctuations in the size of the simulation cell in 
the $xy$ plane ($L_x$ and $L_y$ lengths). This means that 
the in-plane area of the graphene sheets changes along a simulation
run, controlled by the external in-plane stress $P_{xy}$ 
(here $P_{xy} = 0$).
Moreover, C atoms can move in the $z$ coordinate (out-of-plane
direction), so that in general any measure of the {\em real} surface
of a graphene sheet at $T > 0$ will give a value larger than
the in-plane area.

\begin{figure}
\vspace{-0.7cm}
\includegraphics[width=8.0cm]{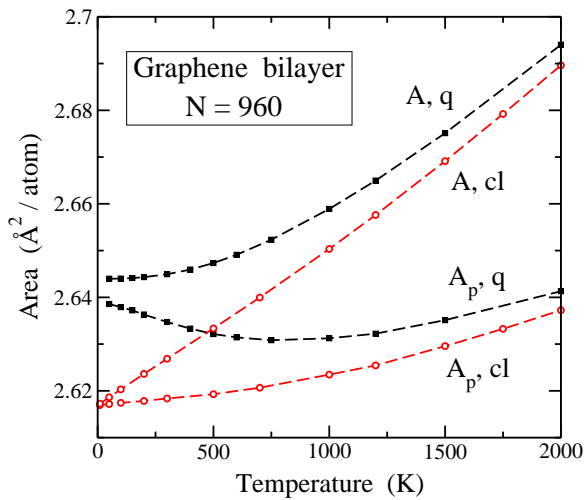}
\vspace{-0.5cm}
\caption{
Mean in-plane ($A_p$) and real area ($A$) as a function
of temperature for graphene bilayers with $N$ = 960.
In both cases open circles and solid squares show results of classical
molecular dynamics (labeled as ``cl'') and quantum PIMD simulations
(labeled as ``q''), respectively.
Error bars are less than the symbol size.
Dashed lines are guides to the eye.
}
\label{f1}
\end{figure}

In Fig.~1 we present the temperature dependence of the in-plane area 
$A_p$ and real area $A$ of a graphene bilayer, as derived from both
classical molecular dynamics (open circles) and PIMD simulations 
(solid squares).
These results were obtained for a simulation cell with $N = 960$.
We discuss first the results of the classical simulations.
For the in-plane area $A_p$, one observes a slow increase at low $T$,
with a slope that becomes larger as the temperature is raised.
For larger simulation cells, the low-temperature slope is found 
to be smaller, and for the largest cells considered here 
($N = 8400$), it is still positive in the whole temperature range.
This contrasts with the results obtained in Ref.~\onlinecite{za10b}
from classical Monte Carlo simulations, where a slight decrease of the
in-plane area was found at low temperature.
This is due to the difference in the interatomic potentials, which 
yields in our case a low-temperature bending constant $\kappa = 1.49$~eV
vs a value of 0.82~eV derived from the earlier version of 
the potential for a graphene sheet.\cite{ra16}
A smaller $\kappa$ causes an enhancement of out-of-plane vibrations, 
which favors a decrease in $A_p$ (see below).

For the classical results, we find that both surfaces $A$ and $A_p$
converge one to the other in the low-$T$ limit, as expected for a 
strictly planar configuration when out-of-plane atomic
displacements vanish as $T \to 0$.
In this classical limit, $A$ and $A_p$ go to a value of 
2.6169 \AA$^2$/atom, which is a little smaller than that obtained for
an isolated graphene sheet, i.e. 2.6190 \AA$^2$/atom.
The real area $A$ derived from classical molecular dynamics 
simulations increases as a function of temperature 
much faster than the in-plane area $A_p$, and
in fact the expansion of $A$ from $T = 0$ to 2000~K is 
a factor of 3.5 larger than that of $A_p$.

In Fig.~1 we also show the temperature dependence of $A$ and $A_p$ 
derived from PIMD simulations.
This dependence is qualitatively different from that obtained in the
classical simulations.  
For the real area we find a behavior similar to that traditionally 
observed for the crystal volume in most 
three-dimensional solids,\cite{ki66}
i.e., a rather flat region at low $T$ and a smooth increase for 
higher temperatures. This is in fact an anharmonic effect, which
in 2D materials such as graphene is combined with out-of-plane
displacements that grow with rising temperature and contribute to
increase the interatomic distance $d_{\rm C-C}$.
At low $T$, $A$ converges to 2.6438 \AA$^2$/atom.
This means a zero-point expansion of $2.5 \times 10^{-2}$ \AA$^2$/atom
with respect to the classical minimum.
This difference between classical and quantum results decreases 
for rising $T$, since nuclear quantum effects become less important.

In contrast to the classical results, the in-plane area derived from 
PIMD simulations decreases for increasing $T$ in a wide temperature 
range and then it increases at higher temperatures after reaching 
a minimum at $T_m \approx 800$~K.
In both cases, classical and quantum, the difference between real
and projected areas increases with temperature. 
In fact, $A_p$ is a measure of a 2D projection of the {\em real} 
graphene surface, and ripples of the actual 
surface have larger amplitudes as temperature is raised.
In the quantum simulations, the in-plane area $A_p$ converges 
at low $T$ to 2.6388 \AA$^2$/atom, 
a value somewhat smaller than that corresponding to the real area.
For $T \to 0$, both derivatives $d A / d T$ and $d A_p / d T$ 
converge to zero, as should happen according to the third law 
of Thermodynamics.  Note that this does not happen in the classical 
approach, a well-known physical inconsistency for this kind of 
calculations at low temperature.\cite{ki66,as76}

There appear two competing effects which explain the temperature 
dependence of $A_p$.
First, the real area $A$ grows as temperature increases in both
cases, classical and quantum.
Second, the presence of ripples in the graphene sheets gives rise 
to a decrease in their 2D projection, i.e., in $A_p$.
At low temperature, the reduction associated to out-of-plane motion 
in the quantum approach predominates over the thermal expansion 
of the actual surface, and consequently we have $d A_p / d T < 0$.
In the classical calculation, however, motion in the $z$ direction
at low temperature is not enough to compensate for the increase in
the area $A$, so that $d A_p / d T > 0$.
At high $T$, the growth of $A$ dominates the decrease 
in projected area caused by out-of-plane atomic motion.
This is in line with the analysis presented by Gao and Huang\cite{ga14}
for the thermal behavior of $A_p$ observed in classical molecular
dynamics simulations of a graphene monolayer, and with the 
calculations by Michel {\em et al.}\cite{mi15b}

\begin{figure}
\vspace{-0.7cm}
\includegraphics[width=8.0cm]{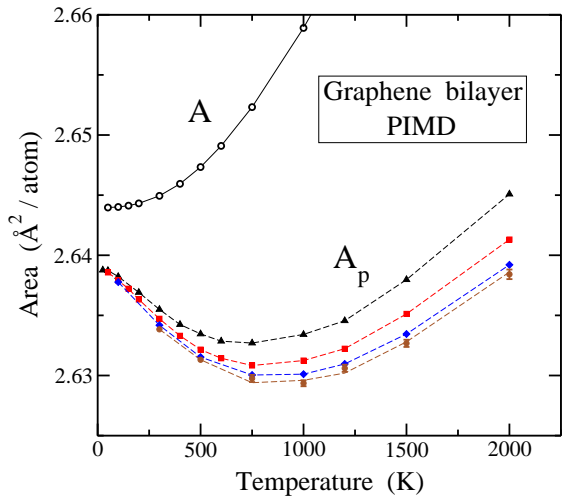}
\vspace{-0.5cm}
\caption{
Mean in-plane ($A_p$) and real area ($A$) vs temperature for graphene
bilayers with several values of $N$, as derived from PIMD simulations.
Data points (solid symbols) for $A_p$ correspond, from top to bottom, to
$N$ = 240 (triangles), 960 (squares), 3840 (diamonds), and
8400 (circles).
Open circles represent results for the real area; data for different cell
sizes are indistinguishable.  Lines are guides to the eye.
Error bars, when not shown, are in the order or less than the symbol size.
}
\label{f2}
\end{figure}

It has been shown earlier from the results of classical and
quantum simulations\cite{he16,ra17,ga14} that the size of the simulation 
cell employed
in the calculations can appreciably affect some magnitudes of
graphene monolayers, such as the in-plane area per atom $A_p$.
In Fig.~2 we present the areas $A$ and $A_p$ vs temperature, as
derived from PIMD simulations of graphene bilayers with several
sizes of the simulation cell.
Data for $A_p$ correspond, from top to bottom, to $N$ = 240 (triangles),
960 (squares), 3840 (diamonds), and 8400 (circles).
The latter size is similar to that employed in earlier classical
Monte Carlo simulations of graphene bilayers by 
Zakharchenko at al.\cite{za10b}

In the results of our quantum simulations one observes that 
$A_p(T)$ displays a minimum in all considered cases. 
This minimum becomes deeper and slightly shifts to higher temperature 
as the system size increases, converging to a value 
$T_m = 850 (\pm 50)$ K for the largest cells discussed here.
For the real area $A$ (open circles) the size effect is very small, 
and in fact it is unobservable at the scale of Fig.~2.
$A$ and $A_p$ become closer to each other as temperature is lowered,
but in the low-$T$ limit $A$ is still larger than $A_p$, and 
we find a difference $A - A_p = 5 \times 10^{-3}$ \AA$^2$/atom 
for $T \to 0$. 
We note that, in spite of the differences in the in-plane area per 
atom for the different system sizes, the results for all system sizes 
converge at low $T$ to a single value.
This is due to the fact that the graphene sheet becomes totally planar
for $T \to 0$ in the classical case and close to planar in the quantum 
model (see above).

\begin{figure}
\vspace{-0.7cm}
\includegraphics[width=8.0cm]{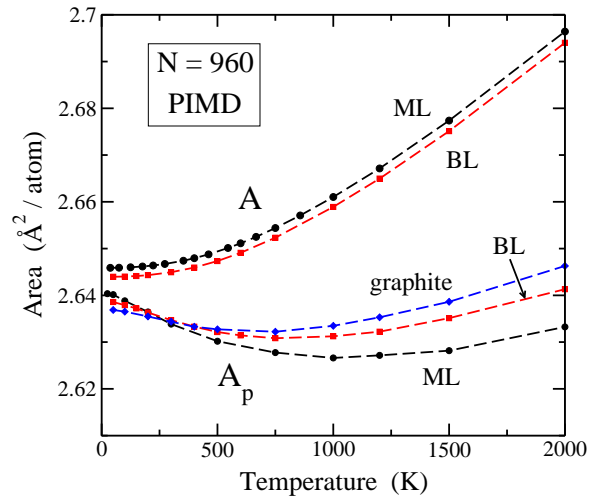}
\vspace{-0.5cm}
\caption{
In-plane projected ($A_p$) and real area ($A$) vs temperature,
as derived from PIMD simulations for monolayer graphene (ML, circles),
bilayer graphene (BL, squares), and graphite (diamonds).
The data shown correspond to $N$ = 960.
Error bars are less than the symbol size.
Dashed lines are guides to the eye.
}
\label{f3}
\end{figure}

It is interesting to compare the thermal behavior of $A$ and $A_p$
for graphene bilayers with that of monolayer graphene.
This is presented in Fig.~3 for the quantum case with $N$ = 960.
For the real area we find a similar behavior for monolayer and bilayer,
with $A$ a little displaced to lower values for the bilayer.
The difference between both cases appears as a rigid shift in the
temperature region shown in the figure up to 2000 K.

The behavior of the in-plane area $A_p$ is somewhat more complex,
since at low $T$, $A_p$ is larger for the monolayer than for 
the bilayer (as the real area $A$), but they become equal 
at $T \approx 200$~K, and at higher temperatures $A_p$ for the bilayer
is clearly larger.
This is a consequence of the competition between the thermal expansion
of the real area and the contraction of the projected in-plane area
associated to out-of-plane atomic displacements, as indicated above.
For the bilayer these displacements (or surface ripples) are smaller 
(see Sec.~VI), and therefore the contraction of $A_p$ is smaller.

We have also plotted in Fig.~3 results for $A_p$ of graphite derived
from PIMD simulations with $N = 960$. These results follow the trend
displayed when passing from the monolayer to the bilayer, i.e.,
at low temperature $A_p$ for graphite is smaller than that corresponding 
to bilayer graphene, and becomes larger than the latter
for $T > 300$~K.
The tendency of the in-plane area of graphite to increase with respect
to the monolayer at low temperature coincides with the results
obtained from density-functional perturbation theory by Mounet
and Marzari.\cite{mo05}
For graphite we obtained a real area $A$ slightly smaller than that
of the bilayer. It converges to 2.6419 \AA$^2$/atom at low temperature
and increases parallel to the real area of the bilayer for rising 
temperature (not shown in Fig.~3 for simplicity).

\begin{figure}
\vspace{-0.7cm}
\includegraphics[width=8.0cm]{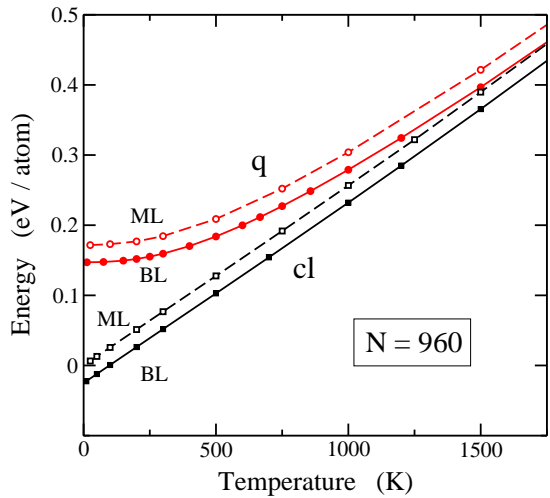}
\vspace{-0.5cm}
\caption{
Temperature dependence of the energy per carbon atom, as derived
from classical (solid squares) and PIMD simulations (solid circles)
of graphene bilayers with $N$ = 960. This corresponds to a
simulation cell including 1920 atoms.
For comparison, open symbols represent data points obtained for
graphene monolayers with $N$ = 960 atoms.
The zero of energy corresponds to the classical minimum for a graphene
monolayer.
Lines are guides to the eye.
Error bars are less than the symbol size.
}
\label{f4}
\end{figure}

\section{Internal energy}

In this section we present and discuss the contributions
to the internal energy of graphene bilayers, corresponding to
our isothermal-isobaric ensemble with $P_{xy} = 0$ and different
temperatures $T$.
In Fig.~4 we show the internal energy (kinetic plus potential
energy) of a graphene bilayer as a function of temperature,
derived from our PIMD simulations for $N = 960$ (solid circles).
For comparison, we also display results of classical simulations for the 
bilayer (solid squares), as well as quantum and classical data for 
a graphene monolayer (open circles and squares).
For the zero of energy we have taken the energy $E_0$ of an isolated 
flat graphene sheet, i.e., the energy minimum in a classical 
calculation at $T = 0$.
One first observes that the zero-temperature classical limit for the 
bilayer is shifted by $E_{\rm int}^0 = -25$ meV/atom, associated to the
stabilization energy due to the interaction between layers. 
This value agrees with that given by Zakharchenko at al.\cite{za10b} 
from classical Monte Carlo simulations of graphene bilayers,
and lies in the intermediate range of binding energies derived from
different {\em ab-initio} calculations for the AB stacking of 
the bilayer.\cite{mo15b}

From PIMD simulations of the bilayer, we find that the internal
energy $E - E_0$ converges at low $T$ to a value of 147 meV/atom, 
which translates into a zero-point energy of 172 meV/atom.
This value is similar to that found for the monolayer
(171 meV/atom), which is not strange since the main contribution to 
the zero-point energy comes from high-frequency in-plane vibrational 
modes, which are not affected by the interaction between layers.
At high temperature, the energy results derived from PIMD simulations 
converge to those of classical simulations, but at $T = 1500$~K we still
observe an appreciable difference between classical and quantum
results for both monolayer and bilayer.

The results shown in Fig.~4 correspond to $N = 960$, as indicated above.
For other cell sizes we obtained results for the internal energy
very close to those shown in the figure, and in fact
indistinguishable from them at the scale of Fig.~4.
This does not happen for other properties of the bilayer,
as indicated below.


\begin{table}[ht]
\caption{Contributions to the internal energy of bilayer graphene
at different temperatures, as derived from PIMD simulations
for $N = 960$. $E_0$ is the energy minimum for an isolated
graphene sheet, taken as reference for the internal energy.
Units of energy are meV/atom. Error bars are in all cases less
than $\pm 0.05$~meV/atom.  }
\vspace{0.3cm}
\centering
\setlength{\tabcolsep}{10pt}
\begin{tabular}{c c c c c}
 $T$ (K) & $E_{\rm int}$  &  $E_{\rm el}$  & $E_{\rm vib}$  &  $E - E_0$
  \\[2mm]
\hline  \\[-2mm]
     50     & -25.0  &  1.5   &  170.8  &  147.3   \\[2mm]

    100     & -25.0  &  1.5   &  171.5  &  148.0   \\[2mm]

    300     & -25.0  &  1.6   &  182.8  &  159.4   \\[2mm]

    500     & -24.9  &  1.9   &  206.9  &  183.9   \\[2mm]

    750     & -24.7  &  2.7   &  249.2  &  227.2   \\[2mm]

   1000     & -24.4  &  3.8   &  299.6  &  279.0   \\[2mm]

   1500     & -23.9  &  7.6   &  413.0  &  396.7   \\[2mm]

   2000     & -23.1  & 13.5   &  534.7  &  525.1   \\[2mm]
\hline  \\[-2mm]
\end {tabular}
\label{tb:coord_seq}
\end{table}

To obtain insight into the origin of the changes in $E$ as a function
of temperature, we may decompose the internal energy in different
contributions as:
\begin{equation}
    E =  E_0 + E_{\rm int} + E_{\rm el} + E_{\rm vib}   \, ,
\label{et}
\end{equation}
where $E_{\rm el}$ is the elastic energy corresponding to an area $A$, 
$E_{\rm vib}$ is the vibrational energy of the system, and
$E_{\rm int}$ is the stabilization energy due to the interaction between
layers. All three contributions change with the temperature.
The expression for the internal energy of the bilayer in Eq.~(\ref{et})
is similar to that corresponding to an isolated
graphene monolayer,\cite{he16} the only difference being the absence
of the term $E_{\rm int}$ in the latter case.
This term corresponds to the binding energy of the graphene layers.
At low temperature $E_{\rm int}$ is indistinguishable from the classical
minimum $E_{\rm int}^0 = -25$~meV/atom, and it slightly changes 
as a function of the interlayer distance, which increases
slowly for rising $T$ (see below).
At the highest temperature considered here ($T = 2000$~K), the interlayer
distance amounts to 3.575 \AA, which gives $E_{\rm int} = -23.1$~meV/atom
(see Table~I).

PIMD simulations directly give $E(T)$, and the elastic energy
can be calculated from the resulting real area $A$.
To obtain the elastic energy corresponding to a given area $A$, we
have calculated the classical energy of a flat graphene sheet with
that area, obtained by isotropically expanding or contracting
the minimum-energy configuration in the layer plane.\cite{he16}
For small changes in the real area, $E_{\rm el}$ is found to change as 
$E_{\rm el}(A) \approx K (A - A_0)^2$, with $K$ = 2.41 eV/\AA$^2$.
At low temperatures most of the energy $E - E_0$ corresponds 
to the vibrational energy, and the 
contribution of the elastic energy increases as the real area grows up.

For a graphene monolayer, the elastic energy $E_{\rm el}$ corresponding
to the area $A$ derived from
PIMD simulations at 300~K amounts to 1.5\% of the internal energy 
$E - E_0$, most of this energy corresponding to the vibrational 
energy $E_{\rm vib}$.
For the graphene bilayer at 300 K, we have $E_{\rm el}$ = 1.6 meV/atom,
to be compared with the internal energy $E - E_0$ = 159.4 meV/atom
(see Table~I).  Then, in this case $E_{\rm el}$ amounts to 1.0\% of the
internal energy.
This value increases to a 2.6\% at 2000 K, since at this temperature
the elastic and internal energies take values of 13.5 and 525.1 meV/atom,
respectively.
We find a reduction of the relative contribution of the elastic
energy to the internal energy of the bilayer, as compared to monolayer
graphene. This is mainly caused by the smaller value of the real area
$A$ for the bilayer (see Fig.~3).

We calculate the vibrational energy $E_{\rm vib}$ from the
results of PIMD simulations by subtracting the elastic and
interaction energy, $E_{\rm el}$ and $E_{\rm int}$, 
from the internal energy $E$ at each temperature [see Eq.~(\ref{et})]. 
At 50 and 300~K, $E_{\rm vib}$ amounts to 171 and 183 meV/atom, 
respectively (see Table~I).
The latter value is somewhat smaller than those found for diamond 
from path-integral simulations at 300 K, namely 195 and 210 meV/atom, 
obtained using Tersoff-type and tight-binding potentials, 
respectively.\cite{he00c,ra06}
For comparison, in a classical harmonic approximation one has: 
$E_{\rm vib}^{\rm cl}(T) = 3 k_B T$ = 77.6 meV/atom.

\begin{figure}
\vspace{-0.7cm}
\includegraphics[width=8.0cm]{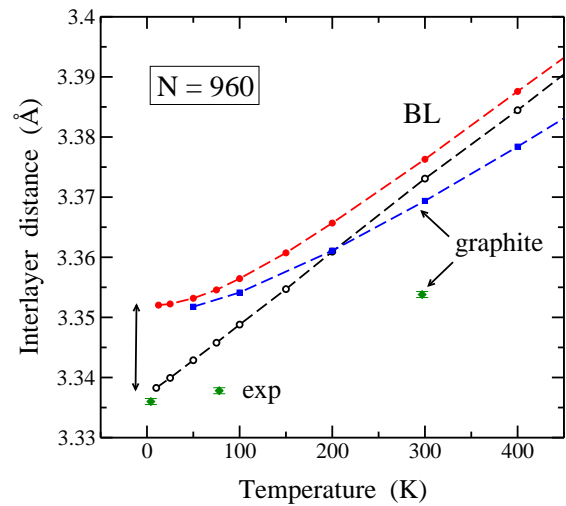}
\vspace{-0.5cm}
\caption{
Mean interlayer distance vs temperature, as derived from
classical (open circles) and PIMD simulations (solid circles)
for graphene bilayers with $N$ = 960.
Solid squares are data points obtained from PIMD simulations
for graphite.
Error bars are smaller than the symbol size.
Dashed lines are guides to the eye.
A vertical arrow indicates the zero-point expansion,
which amounts to 0.015 \AA.
Diamonds (labeled ``exp'') represent data for graphite
derived from x-ray diffraction experiments.\cite{ba55}
}
\label{f5}
\end{figure}

\section{Interlayer spacing and compressibility}

In this section we study the interlayer distance, $c$, and its
fluctuations in bilayer graphene.
In Fig.~5 we present the temperature dependence of the mean equilibrium 
distance $\langle c \rangle$, as derived from our PIMD simulations 
(solid circles). For comparison, we also show the results of 
classical molecular dynamics simulations (open circles).
The classical data converge at low $T$ to an interlayer spacing
$c_0$ = 3.3372~\AA, which corresponds to the minimum-energy
configuration, i.e., totally planar graphene sheets in AB stacking.
These classical calculations yield a linear increase of the mean 
interlayer distance $\langle c \rangle$ for rising temperature, 
with a derivative
$d \langle c \rangle / d T = 1.2 \times 10^{-4}$~\AA/K.
This linear interlayer expansion is similar to that found for lattice 
parameters of crystalline solids in a classical approximation, 
which is known to violate the third law of thermodynamics at
low temperature,\cite{ca60}
since thermal expansion coefficients should vanish for $T \to 0$.
This anomaly of the classical model is remedied in the quantum 
simulations, which yield a vanishing derivative 
$d \langle c \rangle / d T$ in the low-temperature limit,
similarly to the behavior of real and in-plane areas (see Sec.~III).

PIMD simulations give an interlayer spacing larger than
classical calculations, mainly due to zero-point motion of 
the C atoms in the quantum model, which detects anharmonicities
in the interatomic potential even at low temperature.
For $T \to 0$ the interlayer distance converges to 3.3521 \AA. 
Thus, we find a zero-point expansion of $1.5 \times 10^{-2}$ \AA, 
i.e., the mean spacing between layers increases by about 0.5\% with 
respect to the classical prediction.
At room temperature ($T$ = 300 K), the difference between classical and
quantum results amounts to $3.2 \times 10^{-3}$ \AA, about
five times less than in the low-temperature limit.
Size effects due to the finite simulation cells are negligible for the
interlayer spacing.  In fact, for a given temperature, we did not 
find any difference between the results for $\langle c \rangle$ 
obtained for the considered cell sizes, i.e., differences 
were in the order of the error bars found for each 
cell size (less than the symbol size in Fig.~5).

In Fig.~5 we also display the interlayer spacing of graphite derived
from PIMD simulations (solid squares). At low temperature these results
converge to a value very close to the mean distance $\langle c \rangle$ 
for the bilayer.  For rising $T$, the mean spacing $\langle c \rangle$ 
for the bilayer becomes progressively
larger than that of graphite, which agrees with larger thermal 
fluctuations of $c$ in the bilayer (see below).
Diamonds in Fig.~5 represent data points derived for
graphite from x-ray diffraction experiments.\cite{ba55}
Comparing these experimental results with those of our PIMD simulations
for graphite, we find that the interatomic potential LCBOPII
employed here overestimates the interlayer spacing by a 0.5\%.

From the analysis of the interlayer spacing and its fluctuations
one can study the compressibility of bilayer graphene in 
the out-of-plane direction.
The isothermal compressibility $\chi_z$ in the $z$ direction is
defined as
\begin{equation}
  \chi_z = - \frac{1}{\langle V \rangle} 
	    \frac{\partial \langle V \rangle}{\partial P_z}  \, ,
\label{chiz}
\end{equation}
where $V = L_x L_y c$ and $P_z$ is a uniaxial stress in the 
$z$ direction.
At a finite temperature $T$, $\chi_z$ of bilayer graphene can be 
conveniently calculated from PIMD simulations with $P_z = 0$ by using 
the fluctuation formula\cite{la80,he08}
\begin{equation}
    \chi_z = \frac{(\Delta V)^2}{k_B T \langle V \rangle}  \, ,
\label{chiz2}
\end{equation}
where the volume mean-square fluctuations due to changes in the 
interlayer spacing $c$ are given by 
$(\Delta V)^2 = L_x^2 L_y^2 (\Delta c)^2$,
with $(\Delta c)^2 = \langle c^2 \rangle - \langle c \rangle^2$.
Thus, we calculate here the isothermal compressibility $\chi_z$
by using the expression
\begin{equation}
   \chi_z = \frac{L_x L_y}{k_B T} 
      \frac{(\Delta c)^2}{\langle c \rangle}   \, .
\label{chiz3}
\end{equation}

\begin{figure}
\vspace{-0.7cm}
\includegraphics[width=8.0cm]{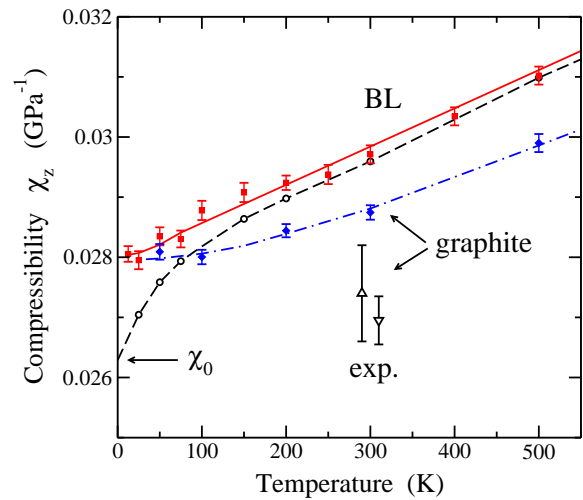}
\vspace{-0.5cm}
\caption{
Compressibiity $\chi_z$ of bilayer graphene as a function of
temperature as derived from PIMD (solid squares) and
classical molecular dynamics simulations (open circles).
Solid diamonds represent results of PIMD simulations for
graphite. Lines are guides to the eye.
Error bars of the classical results are in the order of
the symbol size.
Open triangles indicate results derived from experimental data
of graphite at room temperature:
triangle up from Ref.~\onlinecite{bl70} and
triangle down from Ref.~\onlinecite{ni72}.
A horizontal arrow indicates the classical zero-temperature
limit $\chi_0$.
}
\label{f6}
\end{figure}

In Fig.~6 we display the dependence of $\chi_z$ upon temperature,
as derived from our PIMD simulations (solid squares).
For comparison we also present data for $\chi_z$ obtained from
classical simulations using Eq.~(\ref{chiz3}) (open symbols). 
The classical compressibility $\chi_0$ for $T \to 0$ 
can be calculated from the dependence of the system energy on the 
interlayer spacing $c$ close to $c_0$. For small variations in
$c$ we have an interaction energy
$E_{\rm int} = E_{\rm int}^0 + k (c - c_0)^2 / 2$, 
with a constant $k$ = 0.093 eV \AA$^{-2}$/atom.
Then, the classical zero-temperature compressibility is given by
$\chi_0 = A_0 / 2 c_0 k$, where $A_0$ is the area per atom
(see Appendix A). We find 
$\chi_0 = 2.63 \times 10^{-12}$ cm$^2$ dyn$^{-1}$ or
0.0263 GPa$^{-1}$.
Our classical results converge at low $T$ to this limit, which
is indicated in Fig.~6 with a horizontal arrow.

The compressibilities derived from PIMD simulations at $T\gtrsim 300$~K
are slightly higher that the classical results, and at lower temperatures
they depart progressively one from the other. At low $T$ the quantum
results converge to a value of $2.79(2) \times 10^{-2}$ GPa$^{-1}$.
This means an appreciable increase in $\chi_0$ of a 6\% with
respect to the classical value $\chi_0$.

In connection with the interlayer coupling,
the main difference between the phonon spectrum of monolayer and
bilayer graphene is the appearance in the latter of the so-called
ZO' vibrational band, which is nearly flat in a region of 
2D ${\bf k}$-space close to the $\Gamma$ point.
The frequency of this band at the $\Gamma$ point 
($\omega_{\rm ZO'}^{\Gamma}$, called here $\omega_0$ for simplicity) 
is related to the coupling constant $k$ as  
$\omega_0 = (k_N / M_{\rm red})^{1/2}$,
with $k_ N = 2 N k$ and the reduced mass $M_{\rm red} = \frac12 N m$ 
($m$: carbon atomic mass), so that 
$\omega_0 = 2 (k / m)^{1/2}$. Using the coupling constant
$k$ = 0.093 eV \AA$^{-2}$/atom we find a frequency
$\omega_0$ = 92 cm$^{-1}$.
This corresponds to the layer-breathing $A_{2g}$ Raman-active mode, 
for which a frequency of 89 cm$^{-1}$ has been reported.\cite{li18}

The interlayer coupling has been studied earlier from classical
Monte Carlo simulations in Ref.~\onlinecite{za10b}.
These authors employed a parameter $\gamma$ to describe the
low-frequency part of the ZO' band, which is given in our
terminology by $\gamma = \rho \, \omega_0^2 / 4$,
$\rho$ being the surface mass density. This gives a value
$\gamma$ = 0.035 eV \AA$^{-4}$, in agreement with the low-temperature
results of Monte Carlo simulations.\cite{za10b}

For comparison with the results for the graphene bilayer, we also
present in Fig.~6 some data points for the compressibility of graphite, 
as derived from PIMD simulations and the fluctuation formulas presented
above. At low $T$ these results converge within error bars to the same 
value as the bilayer compressibility, since in both cases the
MSD $(\Delta c)^2$ are nearly identical.
At higher temperatures, however, $(\Delta c)^2$ is smaller for 
graphite, so that its compressibility is lower than that of bilayer
graphene.

It is important to note that the compressibility $\chi_z$ defined
here coincides in the case of graphite with the inverse of the elastic 
constant $C_{33}$ of this material, as this constant relates
strain and stress in the $z$ direction.
In Fig.~6 we present the inverse elastic constant $1 / C_{33}$
determined for pyrolytic graphite from ultrasonic test 
methods\cite{bl70} (triangle up) and from neutron diffraction 
results combined with a force model\cite{ni72} (triangle down).
These data, obtained at room temperature, are slightly displaced 
horizontally around 300 K for clarity.
The compressibility $\chi_z$ for graphite is overestimated by our
simulation results about a 5\% with respect to these data
derived form experiment.
For natural graphite, Komatsu\cite{ko64} obtained a low-temperature 
value $C_{33} = 3.56 \times 10^{11}$ dyn/cm$^2$ from specific-heat 
measurements, which corresponds to $\chi_z$ = 0.0282 GPa$^{-1}$. 
This value (not shown in the figure)
is closer to our results for graphite, but there is no available 
error bar for it.

\section{Out-of-plane motion}

Graphene, as a 2D material in three-dimensional space, has peculiar
properties due to out-of-plane motion, such as the decrease in 
in-plane area $A_p$ for rising temperature discussed in
Sec.~III. For monolayer graphene, motion in the $z$ direction has
been discussed earlier as a function of temperature, applied 
stress, and system size.\cite{ra16,ra17,ga14} 
In this context, one expects that
nuclear quantum effects should be important at relatively low
temperatures, due to the low frequency of vibrational modes in
the $z$ direction, especially those corresponding to
the ZA flexural band. These effects have been studied earlier for 
a graphene monolayer, and in particular the competition between
classical thermal motion and quantum delocalization. In that case
it was found that the outcome of this competition is not trivial,
as it depends significantly not only on temperature, but also
on the applied stress and system size.\cite{he16,he17}

In this section we analyze the mean-square displacements of C atoms 
in the out-of-plane direction, as obtained from PIMD simulations
of bilayer graphene.
We will mainly focus on the character of these atomic displacements,
in order to find whether they can be well explained by a classical model, 
or the carbon atoms noticeably behave as quantum particles. 
We put particular emphasis on the question whether the system size $N$
plays or not a relevant role in this problem.
In this line, PIMD simulations allow us to split vibrational amplitudes 
or atomic delocalization into two parts:
one component associated to thermal (classical-like) motion and 
another corresponding to a proper quantum contribution,
which can be quantified by the mean size of the quantum paths
at a given temperature (see Sec.~II.C).

We will first present results for the MSD obtained in classical 
simulations, where the statistics can be more readily improved
because less computational resources are required. 
Quantum effects are expected to be noticeable mainly
for $T$ lower than room temperature, due to the relatively 
low frequency of the vibrational modes that give the main contribution
to the MSD in the out-of-plane direction.
Note that this does not happen for in-plane modes, with an average
frequency larger than out-of-plane vibrations, so that their
associated zero-point energy is clearly larger.

\begin{figure}
\vspace{-0.7cm}
\includegraphics[width=8.0cm]{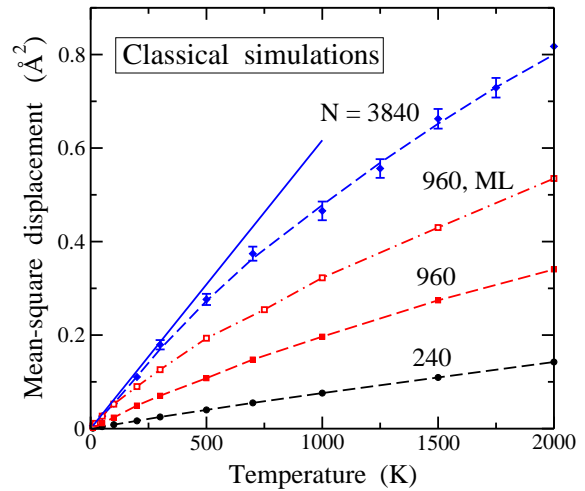}
\vspace{-0.5cm}
\caption{
Temperature dependence of the MSD along the out-of-plane direction,
$(\Delta z)^2$, as derived from classical
molecular dynamics  simulations. Solid symbols represent results for
graphene bilayers with different cell sizes: $N$ = 240 (circles),
960 (squares), and 3840 (diamonds).
Open squares are data points obtained for a graphene monolayer
with $N$ = 960 (labeled as ``ML'').
Error bars, when not shown, are in the order of the symbol size.
Dashed lines are guides to the eye.
A solid line shows the classical MSD for $N = 3840$, obtained
from the harmonic approximation described in the text.
}
\label{f7}
\end{figure}

In Fig.~7 we display results for the MSD in the $z$ direction for 
C atoms in bilayer graphene, as derived for classical molecular 
dynamics simulations for three system sizes (solid symbols). 
From top to bottom: $N$ = 3840 (diamonds), 960 (squares), and
240 (circles). For comparison we also show results for monolayer
graphene with $N$ = 960 (open squares, labeled as ``ML'').
The vibrational amplitude in the $z$ direction increases
with system size, and displays an appreciable anharmonic effect,
since the data obtained from the simulations largely depart from 
linearity. In fact, in a classical harmonic approximation one expects 
for a given system size a MSD increasing linearly with temperature, 
as shown by the solid line corresponding to $N = 3840$.
From the results for monolayer and bilayer with $N = 960$, we
observe an appreciable decrease in the MSD for the bilayer, as
compared with the monolayer. 
The ratio between MSD of monolayer and bilayer 
is 2.3 at 50 K and decreases for rising $T$, taking a value of 
$\approx 1.6$ in the region between 1000 and 2000~K.
The main difference between monolayer and bilayer in this respect is
the less relative importance of ZA modes in the bilayer, due to
the appearance in this case of ZO' and two ZO vibrational bands.

\begin{figure}
\vspace{-0.7cm}
\includegraphics[width=8.0cm]{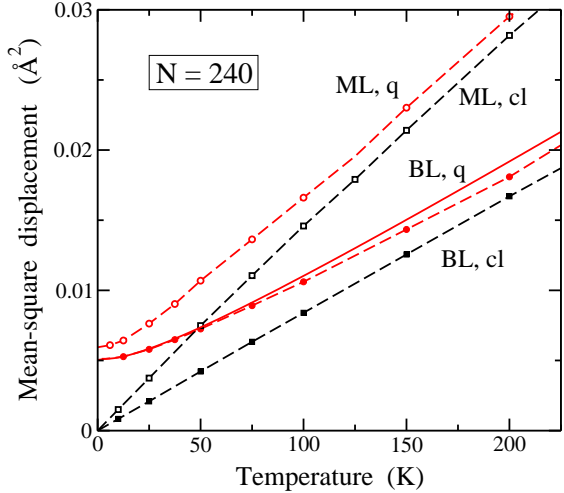}
\vspace{-0.5cm}
\caption{
Mean-square displacement $(\Delta z)^2$ along the out-of-plane
direction for graphene monolayers (ML, open symbols) and bilayers
(BL, solid symbols) with $N$ = 240.
Shown are data obtained from classical (squares, labeled as ``cl'')
and PIMD simulations (circles, labeled as ``q'')
Error bars are in the order of the symbol size.
Dashed lines are guides to the eye.
The solid line represents $(\Delta z)^2$ derived from the quantum HA
described in the text.
}
\label{f8}
\end{figure}

Let us now turn to the results of our quantum simulations.
In Fig.~8 we display data for the motion in the out-of-plane direction, 
corresponding to a cell with $N$ = 240.  Shown are data for the MSD 
$(\Delta z)^2$ obtained for graphene monolayers (open symbols) and 
bilayers (solid symbols).  In both cases we present results from 
classical (squares, labeled as ``cl'') and PIMD simulations
(circles, labeled as ``q'').
At high $T$, classical and quantum data converge one to the other 
in both cases, molayer and bilayer graphene.
One observes that the MSD derived from classical simulations goes  
to zero in the low-temperature limit in each case, whereas PIMD
simulations yield a finite MSD caused by zero-point motion.
We find $(\Delta z)^2_0$ = 5.9 and $4.9 \times 10^{-3}$ \AA$^2$
for the monolayer and bilayer, respectively.
The lower zero-point vibrational amplitude for the bilayer is in line
with an average increase in the frequency of out-of-plane modes
caused by interlayer interactions (less relative importance of the ZA
phonon band).
Comparing the quantum results for monolayer and bilayer at different 
temperatures, we find a ratio between MSD of monolayer and bilayer
which increases from 1.2 at low $T$ to 1.6 at high
temperature, as in the classical limit.
For comparison with the data found from our simulations, we also
present in Fig.~8 the result of a quantum HA (solid line), as derived
from Eq.~(\ref{qz2b}).
At low temperature, the results of HA and PIMD simulations coincide, 
and progressively depart one from the other as temperature is raised. 

The image displayed in Fig.~8 for the atom displacements in the
$z$ direction is qualitatively similar for different system
sizes $N$. The main difference appears in the relative contribution of 
$Q_z^2$ and $C_z^2$ to the total MSD $(\Delta z)^2$.
This is caused by the enhancement of the classical MSD for increasing $N$, 
a fact observed also in data derived from classical molecular dynamics 
simulations of graphene monolayers.\cite{ga14,ra16}

\begin{figure}
\vspace{-0.7cm}
\includegraphics[width=8.0cm]{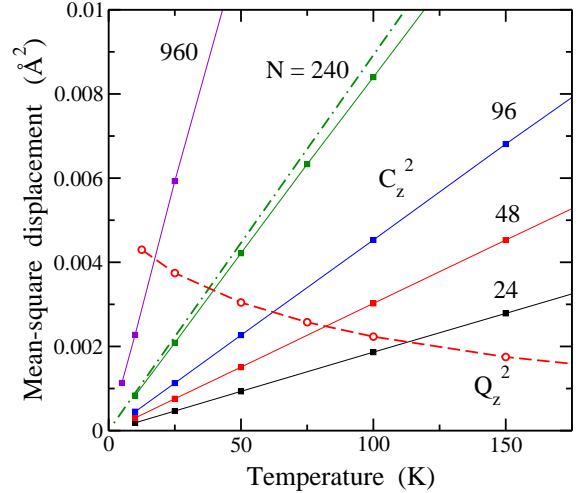}
\vspace{-0.5cm}
\caption{
Mean-square displacement along the out-of-plane direction for graphene
bilayers with different cell sizes, as derived from PIMD simulations.
Solid squares represent results for $C_z^2$ for sizes
$N$ = 960, 240, 96, 48, and 24. Open circles correspond to $Q_z^2$,
where results for different cell sizes are indistinguishable.
Lines are guides to the eye.
Error bars are in the order or less than the symbol size.
The dashed-dotted line displays $C_z^2$ obtained from the HA for
$N = 240$.
}
\label{f9}
\end{figure}

As noted above, an interesting point that can be studied from PIMD 
simulations is the competition between classical-like and
quantum motion as a function of temperature and system size.
One expects that quantum motion should dominate at relatively
low temperatures, but this turns out to be highly dependent
on the size $N$.
In Fig.~9 we present results for $C_z^2$ and $Q_z^2$ as a function
of temperature for several system sizes, from $N$ = 24 to 960.
Data points for the quantum delocalization $Q_z^2$ for different
cell sizes are indistinghuisable at the scale of the figure.
Clear size effects are only found for very small simulation cells
($N < 20$).
On the contrary, the results for $C_z^2$ change very much with
system size. 
In the temperature region shown in Fig.~9 the classical MSD
$C_z^2$ is nearly linear with $T$ for the considered system sizes.
The values of $C_z^2$ presented here, corresponding to the MSD
of the path centroids in PIMD simulations, coincide within error
bars with the atomic MSD obtained from classical simulations.
Note that at higher $T$, $C_z^2$ clearly departs from linearity,
as shown in Fig.~7.
For comparison we also show in Fig.~9 results of the classical HA
given by Eq.~(\ref{cz2b}) for $N = 240$ (dashed-dotted line).
From the data presented in this figure,
it is also worthwhile noting that for system sizes $N < 1000$, 
$C_z^2$ obtained from the simulations scales as $N^{\epsilon}$, 
with an exponent $\epsilon = 0.67$.

Given the increasing slope $d C_z^2 / d T$ for rising $N$,
the crossing of the curves corresponding to $C_z^2$ and $Q_z^2$
moves to lower temperatures.
This means that for large $N$, classical-like motion becomes 
dominant over quantum delocalization for displacemtns in the
$z$ direction.
The quantum contribution $Q_z^2$ converges for $T \to 0$ to
the value given above $(\Delta z)^2_0 = 5 \times 10^{-3}$ \AA$^2$
for the zero-point delocalization in the $z$ direction.
The nuclear quantum delocalization may be estimated from the mean
extension of the quantum paths, i.e., from $Q_z^2$.
At 12.5, 50, and 300~K we find an average extension
$(\Delta z)_Q = (Q_z^2)^{1/2}$  = 0.066, 0.055, and 0.032 \AA,
respectively.

\begin{figure}
\vspace{-0.7cm}
\includegraphics[width=8.0cm]{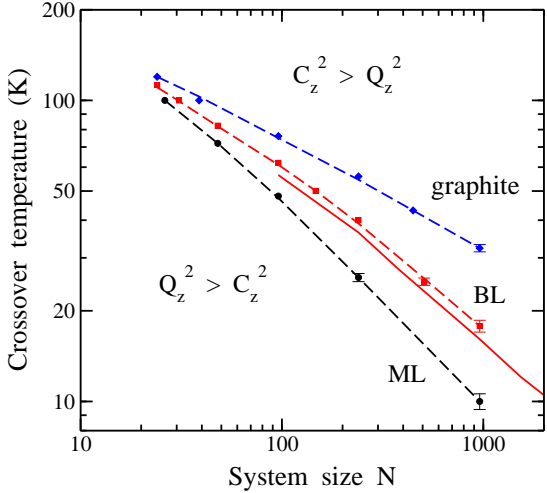}
\vspace{-0.5cm}
\caption{
$N - T$ plane showing the crossover from the region dominated by
quantum delocalization ($Q_z^2 > C_z^2$, below the lines) to
the region dominated by classical-like motion ($C_z^2 > Q_z^2$,
above the lines).
Data points were obtained from PIMD simulations for several system sizes.
From top to bottom: graphite (diamonds), graphene bilayer
(squares), and monolayer (circles).
Error bars, when not shown, are in the order of the symbol size.
The solid line corresponds to the crossover as obtained for the bilayer
from the harmonic approximation presented in the text.
}
\label{f10}
\end{figure}

For a system size $N$, the ratio $Q_z^2 / C_z^2$ decreases for
increasing $T$, and there appears a crossover temperature $T_c$
for which this ratio equals unity.
This corresponds to the crossing of the lines of classical and
quantum displacements, $C_z^2$ and $Q_z^2$, in Fig.~9.
For temperatures $T > T_c$ classical-like motion is dominant for
the atomic motion in the $z$ direction.
In Fig.~10 we display $T_c$ as a function of the system size $N$,
as calculated from PIMD simulations of bilayer graphene.
For comparison, we also present data for a graphene monolayer, as well
as for graphite.  Symbols are results derived from our simulations 
and dashed lines are guides to the eye.
For a given system size $N$, the crossover temperature for the
bilayer is higher than that corresponding to monolayer graphene,
and lower than $T_c$ for graphite. This means that the quantum 
behavior of out-of-plane motion is more relevant for the bilayer
than for an isolated monolayer. It is even more aprreciable for
graphite.   For small $N$ we find similar values of $T_c$ in 
the three cases, and the difference between them increases 
as the system size is raised.
For $N = 960$, we have $T_c$ = $10.1(\pm 0.6)$ K for the monolayer,
$17.8(\pm 0.8)$ K for the bilayer, and $32.3(\pm 0.9)$ K for graphite.
Thus, for this system size, $T_c$ for the bilayer is a factor 1.8 higher 
than for the monolayer.

On the other side, for a given temperature $T$ the ratio 
$C_z^2 / Q_z^2$ grows for increasing $N$, and there is a
system size $N_c$ for which classical motion along the $z$ direction
becomes dominant over quantum delocalization.  
The origin of this behavior is the following.
For a temperature $T$, vibrational modes with frequency
$\omega < \omega_c(T) = k_B T / \hbar$ can be considered
in the classical regime.
The main contribution to $C_z^2$ and $Q_z^2$ comes from ZA flexural 
modes, whose frequency behaves for small $k$ as 
$\rho \, \omega_{\rm ZA}^2 = \sigma k^2 + \kappa k^4$, and
$\omega_{\rm ZA}$ decreases as $k$ is reduced
($\rho$, surface mass density; $\sigma$, effective stress;
$\kappa$, bending modulus).\cite{ra16}
For a size $N$ we have an effective minimum wavevector
$k_0 = 2 \pi / \lambda_{max} \approx 2 \pi / (N A_p)^{1/2}$.
Then, increasing $N$ we attain a size $N_c(T)$ for which additional
increase in size introduces new modes (ZA modes in particular) 
with frequency $\omega < \omega_c$, contributing to rise $C_z^2$ 
more than $Q_z^2$.
Hence, for any $T > 0$ classical-like motion dominates over 
quantum delocalization, when the system size is larger than 
the corresponding $N_c$.

For $N > 100$ the data points corresponding to the bilayer in Fig.~10
can be fitted according to a power-law dependence of the crossover
temperature on system size, i..e, $T_c = b \, N^{-a}$ with a prefactor
$b = 735$ K and an exponent $a = 0.54$.
Extrapolating this dependence to low temperatures,
we have for $T$ = 1~K and 0.1~K crossover sizes
$N_c \approx 2 \times 10^5$ and $1.4 \times 10^7$, respectively.
These sizes are much larger than those actually manageable
in our simulations at these temperatures.

The solid line in Fig.~10 corresponds to the HA model which takes into
account the four vibrational bands of bilayer graphene in the
$z$ direction. For $N \gtrsim 100$, this model predicts a dependence
of the crossover temperature $T_c$ upon $N$ similar to that derived
from PIMD simulations. It yields $T_c$ values smaller than the
simulations, but for large system size we find a power law dependence
with the same exponent in both cases, i.e., $T_c \sim N^{-a}$
with $a = 0.54(\pm 0.02)$.
The main limitations of the HA discussed here are 
the neglect of anharmonicity in the out-of-plane vibrational modes 
(expected to be reasonably small at low $T$) and coupling with 
in-plane modes, which is also expected to increase for rising $T$. 
Taking into account these limitations, the harmonic approach is still 
able to reproduce qualitatively, and nearly quantitatively, the main
features of the competition between classical-like and quantum dynamics
of C atoms in the out-of-plane motion.

\section{Summary}

We have presented results of PIMD simulations of graphene bilayers 
at zero in-plane stress in a wide range of temperatures.
The importance of nuclear quantum effects has been assessed 
by comparing the results of these simulations with those 
obtained from classical molecular dynamics  simulations.
Structural variables such as interatomic distances, as well as
in-plane and real areas are found to increase when quantum nuclear 
motion is considered, even at temperatures in the order of 1000 K.
Zero-point expansion of the graphene layers amounts to about 
1\% of the areas $A$ and $A_p$. This is much larger than the precision
currently reached by diffraction techniques in determining structural
parameters.   The characteristic behavior of the in-plane area $A_p$ 
(decreasing at low $T$ and increasing at high $T$) is a result 
of the coupling between in-plane and out-of-plane modes.
Changes in $A_p$ are , however, less than those obtained for 
isolated graphene layers, as a consequence of interlayer interactions.

We have put particular emphasis on the vibrational motion in the
out-of-plane direction.
Quantum effects appear in these vibrational modes at low temperatures, 
but thermal classical motion becomes dominant for large system size.
This important size effect is a consequence of the enhancement of 
classical-like displacements, whereas quantum delocalization is
nearly unaffected by the system size.
As a result, the crossover temperature $T_c$ at which classical
motion becomes dominant scales for large size as $T_c \sim N^{-a}$
with an exponent $a = 0.54$.

Anharmonicity of the vibrational modes in the $z$ direction 
is clearly observable in the MSDs presented in Figs. 7, 8, and 9.
This affects to both classical and quantum results.
Such anharmonicity shows up markedly in the temperature dependence 
of the in-plane area $A_p$. The real area $A$ is basically
determined by the interatomic distance $d_{\rm C-C}$, 
so its change with temperature is caused by anharmonicity of 
the in-plane modes.   In this context, it is noteworthy that 
a pure harmonic approximation yields rather
accurate results for MSDs at low temperatures, once the vibrational 
frequencies have been calculated for the classical equilibrium 
geometry of the bilayer at $T = 0$.

The compressibility $\chi_z$ of graphene bilayers in the out-of-plane
direction has been obtained from the fluctuations of the interlayer
distance. This method has turned out to be accurate enough to
follow the increase in $\chi_z$ for rising $T$, and also to assess 
the importance of quantum effects at low temperature.
For $T \to 0$ we find an increase in $\chi_z$ of a 6\% with respect
to the classical limit.

A further check of our finite-temperature results would consist in
studying structural and thermodynamic properties of graphene multilayers
from an {\em ab-initio} method.
This is, however, not yet possible taking into account the relatively 
large size of the supercells required to study these properties and
the length of the trajectories necessary for a low statistical uncertainty.

Path-integral simulations analogous to those presented in this paper
could contribute to understand structural and dynamical properties 
of light-atom monolayers on graphene. This is the case of graphane,
where nontrivial quantum features can appear at low temperature,
associated to the small mass of hydrogen.

\begin{acknowledgments}
The authors acknowledge the help of J. H. Los in the implementation 
of the LCBOPII potential.
This work was supported by 
Ministerio de Ciencia, Innovación y Universidades (Spain) through 
Grant FIS2015-64222-C2.
\end{acknowledgments}

\vspace{1cm}

\appendix

\section{Compressibility}

For an interlayer distance $c$ close to the minimum-energy distance
$c_0$, the interaction energy per atom in bilayer graphene
can be written as
\begin{equation}
    E_{\rm int} = E_{\rm int}^0 + \frac12 k (c - c_0)^2
\label{ec}
\end{equation}
with $k$ = 0.093 eV/\AA$^2$.
The classical zero-temperature compressibility in the $z$ direction
is given by
\begin{equation}
 \chi_0 = - \frac{1}{V_0} \left(\frac{\partial V}{\partial P_z} \right)_0 =
        \frac{1}{N A_0 c_0}  
        \left( \frac{\partial^2 \bar{E}_{\rm int}}{\partial V^2} 
	\right)^{-1}_0  \, ,
\label{chi}
\end{equation}
where $P_z$ is a uniaxial stress in the $z$-direction,
$\bar{E}_{\rm int}$ is the interaction energy per simulation cell, 
$\bar{E}_{\rm int} = 2 N E_{\rm int}$,
and changes in $V = N A_p c$ are associated to the variation of 
interlayer spacing $c$.
We find 
\begin{equation}
   \chi_0 = \frac{A_0}{2 \, c_0 \, k}  \, ,
\label{chi2}
\end{equation}
where $A_0$ is the area per atom at $T = 0$ (minimum-energy
configuration).

The isothermal compressibility $\chi_z$ at temperature $T$ can be calculated
from the fluctuation formula:
\begin{equation}
  \chi_z = \frac{(\Delta V)^2}{k_B T \langle V \rangle}  \, ,
\label{chi3}
\end{equation}
with the volume mean-square fluctuations 
$(\Delta V)^2 = N^2 A_p^2 (\Delta c)^2$.

In a harmonic approximation,
the MSD of the interlayer spacing, $(\Delta c)^2$, can be calculated 
from the cost of energy associated to changes of $c$ for a size $N$
($2N$ atoms in bilayer graphene):
$\bar{E}_{\rm int} = \bar{E}_{\rm int}^0 + k N (c - c_0)^2$. 
This yields at temperature $T$ a classical harmonic MSD:
\begin{equation}
    (\Delta c)^2 = \frac{k_B T}{2 N k}
\label{comp}
\end{equation}
which introduced into Eq.~(\ref{chi3}) gives $\chi_z = A_p / 2 c k$.
This expression for the compressibility is valid for interlayer 
distances close to $c_0$,
where $\bar{E}_{\rm int}$ changes as $(c - c_0)^2$.
In general, $\chi_z$ is affected by anharmonicity and has different
values for classical and quantum calculations. The results presented in
Sec.~V were obtained from our simulations by using the fluctuation
formula given in Eq.~(\ref{chi3}).



\begin{thebibliography}{90}
\expandafter\ifx\csname natexlab\endcsname\relax\def\natexlab#1{#1}\fi
\expandafter\ifx\csname bibnamefont\endcsname\relax
  \def\bibnamefont#1{#1}\fi
\expandafter\ifx\csname bibfnamefont\endcsname\relax
  \def\bibfnamefont#1{#1}\fi
\expandafter\ifx\csname citenamefont\endcsname\relax
  \def\citenamefont#1{#1}\fi
\expandafter\ifx\csname url\endcsname\relax
  \def\url#1{\texttt{#1}}\fi
\expandafter\ifx\csname urlprefix\endcsname\relax\def\urlprefix{URL }\fi
\providecommand{\bibinfo}[2]{#2}
\providecommand{\eprint}[2][]{\url{#2}}

\bibitem[{\citenamefont{Geim and Novoselov}(2007)}]{ge07}
\bibinfo{author}{\bibfnamefont{A.~K.} \bibnamefont{Geim}} \bibnamefont{and}
  \bibinfo{author}{\bibfnamefont{K.~S.} \bibnamefont{Novoselov}},
  \bibinfo{journal}{Nature Mater.} \textbf{\bibinfo{volume}{6}},
  \bibinfo{pages}{183} (\bibinfo{year}{2007}).

\bibitem[{\citenamefont{Castro~Neto et~al.}(2009)\citenamefont{Castro~Neto,
  Guinea, Peres, Novoselov, and Geim}}]{ca09b}
\bibinfo{author}{\bibfnamefont{A.~H.} \bibnamefont{Castro~Neto}},
  \bibinfo{author}{\bibfnamefont{F.}~\bibnamefont{Guinea}},
  \bibinfo{author}{\bibfnamefont{N.~M.~R.} \bibnamefont{Peres}},
  \bibinfo{author}{\bibfnamefont{K.~S.} \bibnamefont{Novoselov}},
  \bibnamefont{and} \bibinfo{author}{\bibfnamefont{A.~K.} \bibnamefont{Geim}},
  \bibinfo{journal}{Rev. Mod. Phys.} \textbf{\bibinfo{volume}{81}},
  \bibinfo{pages}{109} (\bibinfo{year}{2009}).

\bibitem[{\citenamefont{Lee et~al.}(2008)\citenamefont{Lee, Wei, Kysar, and
  Hone}}]{le08}
\bibinfo{author}{\bibfnamefont{C.}~\bibnamefont{Lee}},
  \bibinfo{author}{\bibfnamefont{X.}~\bibnamefont{Wei}},
  \bibinfo{author}{\bibfnamefont{J.~W.} \bibnamefont{Kysar}}, \bibnamefont{and}
  \bibinfo{author}{\bibfnamefont{J.}~\bibnamefont{Hone}},
  \bibinfo{journal}{Science} \textbf{\bibinfo{volume}{321}},
  \bibinfo{pages}{385} (\bibinfo{year}{2008}).

\bibitem[{\citenamefont{Gao et~al.}({2018})\citenamefont{Gao, Cao, Cellini,
  Berger, de~Heer, Tosatti, Riedo, and Bongiorno}}]{ga18}
\bibinfo{author}{\bibfnamefont{Y.}~\bibnamefont{Gao}},
  \bibinfo{author}{\bibfnamefont{T.}~\bibnamefont{Cao}},
  \bibinfo{author}{\bibfnamefont{F.}~\bibnamefont{Cellini}},
  \bibinfo{author}{\bibfnamefont{C.}~\bibnamefont{Berger}},
  \bibinfo{author}{\bibfnamefont{W.~A.} \bibnamefont{de~Heer}},
  \bibinfo{author}{\bibfnamefont{E.}~\bibnamefont{Tosatti}},
  \bibinfo{author}{\bibfnamefont{E.}~\bibnamefont{Riedo}}, \bibnamefont{and}
  \bibinfo{author}{\bibfnamefont{A.}~\bibnamefont{Bongiorno}},
  \bibinfo{journal}{Nature Nano.} \textbf{\bibinfo{volume}{{13}}},
  \bibinfo{pages}{{133}} (\bibinfo{year}{{2018}}).

\bibitem[{\citenamefont{Ghosh et~al.}(2008)\citenamefont{Ghosh, Calizo,
  Teweldebrhan, Pokatilov, Nika, Balandin, Bao, Miao, and Lau}}]{gh08b}
\bibinfo{author}{\bibfnamefont{S.}~\bibnamefont{Ghosh}},
  \bibinfo{author}{\bibfnamefont{I.}~\bibnamefont{Calizo}},
  \bibinfo{author}{\bibfnamefont{D.}~\bibnamefont{Teweldebrhan}},
  \bibinfo{author}{\bibfnamefont{E.~P.} \bibnamefont{Pokatilov}},
  \bibinfo{author}{\bibfnamefont{D.~L.} \bibnamefont{Nika}},
  \bibinfo{author}{\bibfnamefont{A.~A.} \bibnamefont{Balandin}},
  \bibinfo{author}{\bibfnamefont{W.}~\bibnamefont{Bao}},
  \bibinfo{author}{\bibfnamefont{F.}~\bibnamefont{Miao}}, \bibnamefont{and}
  \bibinfo{author}{\bibfnamefont{C.~N.} \bibnamefont{Lau}},
  \bibinfo{journal}{Appl. Phys. Lett.} \textbf{\bibinfo{volume}{92}},
  \bibinfo{pages}{151911} (\bibinfo{year}{2008}).

\bibitem[{\citenamefont{Seol et~al.}(2010)\citenamefont{Seol, Jo, Moore,
  Lindsay, Aitken, Pettes, Li, Yao, Huang, Broido et~al.}}]{se10}
\bibinfo{author}{\bibfnamefont{J.~H.} \bibnamefont{Seol}},
  \bibinfo{author}{\bibfnamefont{I.}~\bibnamefont{Jo}},
  \bibinfo{author}{\bibfnamefont{A.~L.} \bibnamefont{Moore}},
  \bibinfo{author}{\bibfnamefont{L.}~\bibnamefont{Lindsay}},
  \bibinfo{author}{\bibfnamefont{Z.~H.} \bibnamefont{Aitken}},
  \bibinfo{author}{\bibfnamefont{M.~T.} \bibnamefont{Pettes}},
  \bibinfo{author}{\bibfnamefont{X.}~\bibnamefont{Li}},
  \bibinfo{author}{\bibfnamefont{Z.}~\bibnamefont{Yao}},
  \bibinfo{author}{\bibfnamefont{R.}~\bibnamefont{Huang}},
  \bibinfo{author}{\bibfnamefont{D.}~\bibnamefont{Broido}},
  \bibnamefont{et~al.}, \bibinfo{journal}{Science}
  \textbf{\bibinfo{volume}{328}}, \bibinfo{pages}{213} (\bibinfo{year}{2010}).

\bibitem[{\citenamefont{Balandin}(2011)}]{ba11}
\bibinfo{author}{\bibfnamefont{A.~A.} \bibnamefont{Balandin}},
  \bibinfo{journal}{Nature Mater.} \textbf{\bibinfo{volume}{10}},
  \bibinfo{pages}{569} (\bibinfo{year}{2011}).

\bibitem[{\citenamefont{Zhang et~al.}(2016)\citenamefont{Zhang, Han, Qiao, Tan,
  Wang, Zhang, and Tan}}]{zh16b}
\bibinfo{author}{\bibfnamefont{X.}~\bibnamefont{Zhang}},
  \bibinfo{author}{\bibfnamefont{W.-P.} \bibnamefont{Han}},
  \bibinfo{author}{\bibfnamefont{X.-F.} \bibnamefont{Qiao}},
  \bibinfo{author}{\bibfnamefont{Q.-H.} \bibnamefont{Tan}},
  \bibinfo{author}{\bibfnamefont{Y.-F.} \bibnamefont{Wang}},
  \bibinfo{author}{\bibfnamefont{J.}~\bibnamefont{Zhang}}, \bibnamefont{and}
  \bibinfo{author}{\bibfnamefont{P.-H.} \bibnamefont{Tan}},
  \bibinfo{journal}{Carbon} \textbf{\bibinfo{volume}{99}}, \bibinfo{pages}{118}
  (\bibinfo{year}{2016}).

\bibitem[{\citenamefont{Novoselov et~al.}(2006)\citenamefont{Novoselov, McCann,
  Morozov, Fal'ko, Katsnelson, Zeitler, Jiang, Schedin, and Geim}}]{no06}
\bibinfo{author}{\bibfnamefont{K.~S.} \bibnamefont{Novoselov}},
  \bibinfo{author}{\bibfnamefont{E.}~\bibnamefont{McCann}},
  \bibinfo{author}{\bibfnamefont{S.~V.} \bibnamefont{Morozov}},
  \bibinfo{author}{\bibfnamefont{V.~I.} \bibnamefont{Fal'ko}},
  \bibinfo{author}{\bibfnamefont{M.~I.} \bibnamefont{Katsnelson}},
  \bibinfo{author}{\bibfnamefont{U.}~\bibnamefont{Zeitler}},
  \bibinfo{author}{\bibfnamefont{D.}~\bibnamefont{Jiang}},
  \bibinfo{author}{\bibfnamefont{F.}~\bibnamefont{Schedin}}, \bibnamefont{and}
  \bibinfo{author}{\bibfnamefont{A.~K.} \bibnamefont{Geim}},
  \bibinfo{journal}{Nature Phys.} \textbf{\bibinfo{volume}{2}},
  \bibinfo{pages}{177} (\bibinfo{year}{2006}).

\bibitem[{\citenamefont{Meyer et~al.}(2007)\citenamefont{Meyer, Geim,
  Katsnelson, Novoselov, Obergfell, Roth, Girit, and Zettl}}]{me07b}
\bibinfo{author}{\bibfnamefont{J.~C.} \bibnamefont{Meyer}},
  \bibinfo{author}{\bibfnamefont{A.~K.} \bibnamefont{Geim}},
  \bibinfo{author}{\bibfnamefont{M.~I.} \bibnamefont{Katsnelson}},
  \bibinfo{author}{\bibfnamefont{K.~S.} \bibnamefont{Novoselov}},
  \bibinfo{author}{\bibfnamefont{D.}~\bibnamefont{Obergfell}},
  \bibinfo{author}{\bibfnamefont{S.}~\bibnamefont{Roth}},
  \bibinfo{author}{\bibfnamefont{C.}~\bibnamefont{Girit}}, \bibnamefont{and}
  \bibinfo{author}{\bibfnamefont{A.}~\bibnamefont{Zettl}},
  \bibinfo{journal}{Solid State Commun.} \textbf{\bibinfo{volume}{143}},
  \bibinfo{pages}{101} (\bibinfo{year}{2007}).

\bibitem[{\citenamefont{Gibertini et~al.}(2010)\citenamefont{Gibertini,
  Tomadin, Polini, Fasolino, and Katsnelson}}]{gi10}
\bibinfo{author}{\bibfnamefont{M.}~\bibnamefont{Gibertini}},
  \bibinfo{author}{\bibfnamefont{A.}~\bibnamefont{Tomadin}},
  \bibinfo{author}{\bibfnamefont{M.}~\bibnamefont{Polini}},
  \bibinfo{author}{\bibfnamefont{A.}~\bibnamefont{Fasolino}}, \bibnamefont{and}
  \bibinfo{author}{\bibfnamefont{M.~I.} \bibnamefont{Katsnelson}},
  \bibinfo{journal}{Phys. Rev. B} \textbf{\bibinfo{volume}{81}},
  \bibinfo{pages}{125437} (\bibinfo{year}{2010}).

\bibitem[{\citenamefont{Cao et~al.}(2018)\citenamefont{Cao, Fatemi, Fang,
  Watanabe, Taniguchi, Kaxiras, and Jarillo-Herrero}}]{ca18}
\bibinfo{author}{\bibfnamefont{Y.}~\bibnamefont{Cao}},
  \bibinfo{author}{\bibfnamefont{V.}~\bibnamefont{Fatemi}},
  \bibinfo{author}{\bibfnamefont{S.}~\bibnamefont{Fang}},
  \bibinfo{author}{\bibfnamefont{K.}~\bibnamefont{Watanabe}},
  \bibinfo{author}{\bibfnamefont{T.}~\bibnamefont{Taniguchi}},
  \bibinfo{author}{\bibfnamefont{E.}~\bibnamefont{Kaxiras}}, \bibnamefont{and}
  \bibinfo{author}{\bibfnamefont{P.}~\bibnamefont{Jarillo-Herrero}},
  \bibinfo{journal}{Nature} \textbf{\bibinfo{volume}{556}}, \bibinfo{pages}{43}
  (\bibinfo{year}{2018}).

\bibitem[{\citenamefont{Cao et~al.}({2018})\citenamefont{Cao, Fatemi, Demir,
  Fang, Tomarken, Luo, Sanchez-Yamagishi, Watanabe, Taniguchi, Kaxiras
  et~al.}}]{ca18b}
\bibinfo{author}{\bibfnamefont{Y.}~\bibnamefont{Cao}},
  \bibinfo{author}{\bibfnamefont{V.}~\bibnamefont{Fatemi}},
  \bibinfo{author}{\bibfnamefont{A.}~\bibnamefont{Demir}},
  \bibinfo{author}{\bibfnamefont{S.}~\bibnamefont{Fang}},
  \bibinfo{author}{\bibfnamefont{S.~L.} \bibnamefont{Tomarken}},
  \bibinfo{author}{\bibfnamefont{J.~Y.} \bibnamefont{Luo}},
  \bibinfo{author}{\bibfnamefont{J.~D.} \bibnamefont{Sanchez-Yamagishi}},
  \bibinfo{author}{\bibfnamefont{K.}~\bibnamefont{Watanabe}},
  \bibinfo{author}{\bibfnamefont{T.}~\bibnamefont{Taniguchi}},
  \bibinfo{author}{\bibfnamefont{E.}~\bibnamefont{Kaxiras}},
  \bibnamefont{et~al.}, \bibinfo{journal}{Nature}
  \textbf{\bibinfo{volume}{{556}}}, \bibinfo{pages}{{80}}
  (\bibinfo{year}{{2018}}).

\bibitem[{\citenamefont{Yndurain}({2019})}]{yn19}
\bibinfo{author}{\bibfnamefont{F.}~\bibnamefont{Yndurain}},
  \bibinfo{journal}{Phys. Rev. B} \textbf{\bibinfo{volume}{{99}}},
  \bibinfo{pages}{{045423}} (\bibinfo{year}{{2019}}).

\bibitem[{\citenamefont{Safran}(1994)}]{sa94}
\bibinfo{author}{\bibfnamefont{S.~A.} \bibnamefont{Safran}},
  \emph{\bibinfo{title}{Statistical Thermodynamics of Surfaces, Interfaces, and
  Membranes}} (\bibinfo{publisher}{Addison Wesley}, \bibinfo{address}{New
  York}, \bibinfo{year}{1994}).

\bibitem[{\citenamefont{Nelson et~al.}(2004)\citenamefont{Nelson, Piran, and
  Weinberg}}]{ne04}
\bibinfo{author}{\bibfnamefont{D.}~\bibnamefont{Nelson}},
  \bibinfo{author}{\bibfnamefont{T.}~\bibnamefont{Piran}}, \bibnamefont{and}
  \bibinfo{author}{\bibfnamefont{S.}~\bibnamefont{Weinberg}},
  \emph{\bibinfo{title}{Statistical Mechanics of Membranes and Surfaces}}
  (\bibinfo{publisher}{World Scientific}, \bibinfo{address}{London},
  \bibinfo{year}{2004}).

\bibitem[{\citenamefont{Chac\'on et~al.}(2015)\citenamefont{Chac\'on, Tarazona,
  and Bresme}}]{ch15}
\bibinfo{author}{\bibfnamefont{E.}~\bibnamefont{Chac\'on}},
  \bibinfo{author}{\bibfnamefont{P.}~\bibnamefont{Tarazona}}, \bibnamefont{and}
  \bibinfo{author}{\bibfnamefont{F.}~\bibnamefont{Bresme}},
  \bibinfo{journal}{J. Chem. Phys.} \textbf{\bibinfo{volume}{143}},
  \bibinfo{pages}{034706} (\bibinfo{year}{2015}).

\bibitem[{\citenamefont{Ruiz-Herrero et~al.}(2012)\citenamefont{Ruiz-Herrero,
  Velasco, and Hagan}}]{ru12}
\bibinfo{author}{\bibfnamefont{T.}~\bibnamefont{Ruiz-Herrero}},
  \bibinfo{author}{\bibfnamefont{E.}~\bibnamefont{Velasco}}, \bibnamefont{and}
  \bibinfo{author}{\bibfnamefont{M.~F.} \bibnamefont{Hagan}},
  \bibinfo{journal}{J. Phys. Chem. B} \textbf{\bibinfo{volume}{116}},
  \bibinfo{pages}{9595} (\bibinfo{year}{2012}).

\bibitem[{\citenamefont{Fasolino et~al.}(2007)\citenamefont{Fasolino, Los, and
  Katsnelson}}]{fa07}
\bibinfo{author}{\bibfnamefont{A.}~\bibnamefont{Fasolino}},
  \bibinfo{author}{\bibfnamefont{J.~H.} \bibnamefont{Los}}, \bibnamefont{and}
  \bibinfo{author}{\bibfnamefont{M.~I.} \bibnamefont{Katsnelson}},
  \bibinfo{journal}{Nature Mater.} \textbf{\bibinfo{volume}{6}},
  \bibinfo{pages}{858} (\bibinfo{year}{2007}).

\bibitem[{\citenamefont{Alofi and Srivastava}(2013)}]{al13}
\bibinfo{author}{\bibfnamefont{A.}~\bibnamefont{Alofi}} \bibnamefont{and}
  \bibinfo{author}{\bibfnamefont{G.~P.} \bibnamefont{Srivastava}},
  \bibinfo{journal}{Phys. Rev. B} \textbf{\bibinfo{volume}{87}},
  \bibinfo{pages}{115421} (\bibinfo{year}{2013}).

\bibitem[{\citenamefont{Amorim et~al.}(2014)\citenamefont{Amorim, Roldan,
  Cappelluti, Fasolino, Guinea, and Katsnelson}}]{am14}
\bibinfo{author}{\bibfnamefont{B.}~\bibnamefont{Amorim}},
  \bibinfo{author}{\bibfnamefont{R.}~\bibnamefont{Roldan}},
  \bibinfo{author}{\bibfnamefont{E.}~\bibnamefont{Cappelluti}},
  \bibinfo{author}{\bibfnamefont{A.}~\bibnamefont{Fasolino}},
  \bibinfo{author}{\bibfnamefont{F.}~\bibnamefont{Guinea}}, \bibnamefont{and}
  \bibinfo{author}{\bibfnamefont{M.~I.} \bibnamefont{Katsnelson}},
  \bibinfo{journal}{Phys. Rev. B} \textbf{\bibinfo{volume}{89}},
  \bibinfo{pages}{224307} (\bibinfo{year}{2014}).

\bibitem[{\citenamefont{Mann et~al.}(2016)\citenamefont{Mann, Rani, Kumar,
  Dubey, and Jindal}}]{ma16}
\bibinfo{author}{\bibfnamefont{S.}~\bibnamefont{Mann}},
  \bibinfo{author}{\bibfnamefont{P.}~\bibnamefont{Rani}},
  \bibinfo{author}{\bibfnamefont{R.}~\bibnamefont{Kumar}},
  \bibinfo{author}{\bibfnamefont{G.~S.} \bibnamefont{Dubey}}, \bibnamefont{and}
  \bibinfo{author}{\bibfnamefont{V.~K.} \bibnamefont{Jindal}},
  \bibinfo{journal}{RSC Adv.} \textbf{\bibinfo{volume}{6}},
  \bibinfo{pages}{12158} (\bibinfo{year}{2016}).

\bibitem[{\citenamefont{Coquand and Mouhanna}(2016)}]{co16}
\bibinfo{author}{\bibfnamefont{O.}~\bibnamefont{Coquand}} \bibnamefont{and}
  \bibinfo{author}{\bibfnamefont{D.}~\bibnamefont{Mouhanna}},
  \bibinfo{journal}{Phys. Rev. E} \textbf{\bibinfo{volume}{94}},
  \bibinfo{pages}{032125} (\bibinfo{year}{2016}).

\bibitem[{\citenamefont{Cadelano et~al.}(2009)\citenamefont{Cadelano, Palla,
  Giordano, and Colombo}}]{ca09c}
\bibinfo{author}{\bibfnamefont{E.}~\bibnamefont{Cadelano}},
  \bibinfo{author}{\bibfnamefont{P.~L.} \bibnamefont{Palla}},
  \bibinfo{author}{\bibfnamefont{S.}~\bibnamefont{Giordano}}, \bibnamefont{and}
  \bibinfo{author}{\bibfnamefont{L.}~\bibnamefont{Colombo}},
  \bibinfo{journal}{Phys. Rev. Lett.} \textbf{\bibinfo{volume}{102}},
  \bibinfo{pages}{235502} (\bibinfo{year}{2009}).

\bibitem[{\citenamefont{Akatyeva and Dumitrica}(2012)}]{ak12}
\bibinfo{author}{\bibfnamefont{E.}~\bibnamefont{Akatyeva}} \bibnamefont{and}
  \bibinfo{author}{\bibfnamefont{T.}~\bibnamefont{Dumitrica}},
  \bibinfo{journal}{J. Chem. Phys.} \textbf{\bibinfo{volume}{137}},
  \bibinfo{pages}{234702} (\bibinfo{year}{2012}).

\bibitem[{\citenamefont{Chechin et~al.}(2014)\citenamefont{Chechin, Dmitriev,
  Lobzenko, and Ryabov}}]{ch14}
\bibinfo{author}{\bibfnamefont{G.~M.} \bibnamefont{Chechin}},
  \bibinfo{author}{\bibfnamefont{S.~V.} \bibnamefont{Dmitriev}},
  \bibinfo{author}{\bibfnamefont{I.~P.} \bibnamefont{Lobzenko}},
  \bibnamefont{and} \bibinfo{author}{\bibfnamefont{D.~S.}
  \bibnamefont{Ryabov}}, \bibinfo{journal}{Phys. Rev. B}
  \textbf{\bibinfo{volume}{90}}, \bibinfo{pages}{045432}
  (\bibinfo{year}{2014}).

\bibitem[{\citenamefont{Magnin et~al.}(2014)\citenamefont{Magnin, Foerster,
  Rabilloud, Calvo, Zappelli, and Bichara}}]{ma14}
\bibinfo{author}{\bibfnamefont{Y.}~\bibnamefont{Magnin}},
  \bibinfo{author}{\bibfnamefont{G.~D.} \bibnamefont{Foerster}},
  \bibinfo{author}{\bibfnamefont{F.}~\bibnamefont{Rabilloud}},
  \bibinfo{author}{\bibfnamefont{F.}~\bibnamefont{Calvo}},
  \bibinfo{author}{\bibfnamefont{A.}~\bibnamefont{Zappelli}}, \bibnamefont{and}
  \bibinfo{author}{\bibfnamefont{C.}~\bibnamefont{Bichara}},
  \bibinfo{journal}{J. Phys.: Condens. Matter} \textbf{\bibinfo{volume}{26}},
  \bibinfo{pages}{185401} (\bibinfo{year}{2014}).

\bibitem[{\citenamefont{Los et~al.}(2016)\citenamefont{Los, Fasolino, and
  Katsnelson}}]{lo16}
\bibinfo{author}{\bibfnamefont{J.~H.} \bibnamefont{Los}},
  \bibinfo{author}{\bibfnamefont{A.}~\bibnamefont{Fasolino}}, \bibnamefont{and}
  \bibinfo{author}{\bibfnamefont{M.~I.} \bibnamefont{Katsnelson}},
  \bibinfo{journal}{Phys. Rev. Lett.} \textbf{\bibinfo{volume}{116}},
  \bibinfo{pages}{015901} (\bibinfo{year}{2016}).

\bibitem[{\citenamefont{Politano et~al.}(2011)\citenamefont{Politano, Borca,
  Minniti, Hinarejos, Vazquez~de Parga, Farias, and Miranda}}]{po11}
\bibinfo{author}{\bibfnamefont{A.}~\bibnamefont{Politano}},
  \bibinfo{author}{\bibfnamefont{B.}~\bibnamefont{Borca}},
  \bibinfo{author}{\bibfnamefont{M.}~\bibnamefont{Minniti}},
  \bibinfo{author}{\bibfnamefont{J.~J.} \bibnamefont{Hinarejos}},
  \bibinfo{author}{\bibfnamefont{A.~L.} \bibnamefont{Vazquez~de Parga}},
  \bibinfo{author}{\bibfnamefont{D.}~\bibnamefont{Farias}}, \bibnamefont{and}
  \bibinfo{author}{\bibfnamefont{R.}~\bibnamefont{Miranda}},
  \bibinfo{journal}{Phys. Rev. B} \textbf{\bibinfo{volume}{84}},
  \bibinfo{pages}{035450} (\bibinfo{year}{2011}).

\bibitem[{\citenamefont{Gillan}(1988)}]{gi88}
\bibinfo{author}{\bibfnamefont{M.~J.} \bibnamefont{Gillan}},
  \bibinfo{journal}{Phil. Mag. A} \textbf{\bibinfo{volume}{58}},
  \bibinfo{pages}{257} (\bibinfo{year}{1988}).

\bibitem[{\citenamefont{Ceperley}(1995)}]{ce95}
\bibinfo{author}{\bibfnamefont{D.~M.} \bibnamefont{Ceperley}},
  \bibinfo{journal}{Rev. Mod. Phys.} \textbf{\bibinfo{volume}{67}},
  \bibinfo{pages}{279} (\bibinfo{year}{1995}).

\bibitem[{\citenamefont{Brito et~al.}(2015)\citenamefont{Brito, C\^andido, Hai,
  and Peeters}}]{br15}
\bibinfo{author}{\bibfnamefont{B.~G.~A.} \bibnamefont{Brito}},
  \bibinfo{author}{\bibfnamefont{L.}~\bibnamefont{C\^andido}},
  \bibinfo{author}{\bibfnamefont{G.-Q.} \bibnamefont{Hai}}, \bibnamefont{and}
  \bibinfo{author}{\bibfnamefont{F.~M.} \bibnamefont{Peeters}},
  \bibinfo{journal}{Phys. Rev. B} \textbf{\bibinfo{volume}{92}},
  \bibinfo{pages}{195416} (\bibinfo{year}{2015}).

\bibitem[{\citenamefont{Herrero and Ram\'irez}(2016)}]{he16}
\bibinfo{author}{\bibfnamefont{C.~P.} \bibnamefont{Herrero}} \bibnamefont{and}
  \bibinfo{author}{\bibfnamefont{R.}~\bibnamefont{Ram\'irez}},
  \bibinfo{journal}{J. Chem. Phys.} \textbf{\bibinfo{volume}{145}},
  \bibinfo{pages}{224701} (\bibinfo{year}{2016}).

\bibitem[{\citenamefont{Herrero and Ram\'irez}(2018{\natexlab{a}})}]{he18}
\bibinfo{author}{\bibfnamefont{C.~P.} \bibnamefont{Herrero}} \bibnamefont{and}
  \bibinfo{author}{\bibfnamefont{R.}~\bibnamefont{Ram\'irez}},
  \bibinfo{journal}{J. Chem. Phys.} \textbf{\bibinfo{volume}{148}},
  \bibinfo{pages}{102302} (\bibinfo{year}{2018}{\natexlab{a}}).

\bibitem[{\citenamefont{Hasik et~al.}({2018})\citenamefont{Hasik, Tosatti, and
  Martonak}}]{ha18}
\bibinfo{author}{\bibfnamefont{J.}~\bibnamefont{Hasik}},
  \bibinfo{author}{\bibfnamefont{E.}~\bibnamefont{Tosatti}}, \bibnamefont{and}
  \bibinfo{author}{\bibfnamefont{R.}~\bibnamefont{Martonak}},
  \bibinfo{journal}{Phys. Rev. B} \textbf{\bibinfo{volume}{{97}}},
  \bibinfo{pages}{{140301}} (\bibinfo{year}{{2018}}).

\bibitem[{\citenamefont{Gao and Huang}(2014)}]{ga14}
\bibinfo{author}{\bibfnamefont{W.}~\bibnamefont{Gao}} \bibnamefont{and}
  \bibinfo{author}{\bibfnamefont{R.}~\bibnamefont{Huang}}, \bibinfo{journal}{J.
  Mech. Phys. Solids} \textbf{\bibinfo{volume}{66}}, \bibinfo{pages}{42}
  (\bibinfo{year}{2014}).

\bibitem[{\citenamefont{Ram\'{\i}rez et~al.}(2006)\citenamefont{Ram\'{\i}rez,
  Herrero, and Hern\'andez}}]{ra06}
\bibinfo{author}{\bibfnamefont{R.}~\bibnamefont{Ram\'{\i}rez}},
  \bibinfo{author}{\bibfnamefont{C.~P.} \bibnamefont{Herrero}},
  \bibnamefont{and} \bibinfo{author}{\bibfnamefont{E.~R.}
  \bibnamefont{Hern\'andez}}, \bibinfo{journal}{Phys. Rev. B}
  \textbf{\bibinfo{volume}{73}}, \bibinfo{pages}{245202}
  (\bibinfo{year}{2006}).

\bibitem[{\citenamefont{Herrero and Ram\'{\i}rez}(2010)}]{he10}
\bibinfo{author}{\bibfnamefont{C.~P.} \bibnamefont{Herrero}} \bibnamefont{and}
  \bibinfo{author}{\bibfnamefont{R.}~\bibnamefont{Ram\'{\i}rez}},
  \bibinfo{journal}{Phys. Rev. B} \textbf{\bibinfo{volume}{82}},
  \bibinfo{pages}{174117} (\bibinfo{year}{2010}).

\bibitem[{\citenamefont{Kwon and Ceperley}(2012)}]{kw12}
\bibinfo{author}{\bibfnamefont{Y.}~\bibnamefont{Kwon}} \bibnamefont{and}
  \bibinfo{author}{\bibfnamefont{D.~M.} \bibnamefont{Ceperley}},
  \bibinfo{journal}{Phys. Rev. B} \textbf{\bibinfo{volume}{85}},
  \bibinfo{pages}{224501} (\bibinfo{year}{2012}).

\bibitem[{\citenamefont{Herrero and Ram\'irez}(2009)}]{he09a}
\bibinfo{author}{\bibfnamefont{C.~P.} \bibnamefont{Herrero}} \bibnamefont{and}
  \bibinfo{author}{\bibfnamefont{R.}~\bibnamefont{Ram\'irez}},
  \bibinfo{journal}{Phys. Rev. B} \textbf{\bibinfo{volume}{79}},
  \bibinfo{pages}{115429} (\bibinfo{year}{2009}).

\bibitem[{\citenamefont{Davidson et~al.}(2014)\citenamefont{Davidson, Klimes,
  Alfe, and Michaelides}}]{da14}
\bibinfo{author}{\bibfnamefont{E.~R.~M.} \bibnamefont{Davidson}},
  \bibinfo{author}{\bibfnamefont{J.}~\bibnamefont{Klimes}},
  \bibinfo{author}{\bibfnamefont{D.}~\bibnamefont{Alfe}}, \bibnamefont{and}
  \bibinfo{author}{\bibfnamefont{A.}~\bibnamefont{Michaelides}},
  \bibinfo{journal}{ACS Nano} \textbf{\bibinfo{volume}{8}},
  \bibinfo{pages}{9905} (\bibinfo{year}{2014}).

\bibitem[{\citenamefont{Feynman}(1972)}]{fe72}
\bibinfo{author}{\bibfnamefont{R.~P.} \bibnamefont{Feynman}},
  \emph{\bibinfo{title}{Statistical Mechanics}}
  (\bibinfo{publisher}{Addison-Wesley}, \bibinfo{address}{New York},
  \bibinfo{year}{1972}).

\bibitem[{\citenamefont{Herrero and Ram\'irez}(2014)}]{he14}
\bibinfo{author}{\bibfnamefont{C.~P.} \bibnamefont{Herrero}} \bibnamefont{and}
  \bibinfo{author}{\bibfnamefont{R.}~\bibnamefont{Ram\'irez}},
  \bibinfo{journal}{J. Phys.: Condens. Matter} \textbf{\bibinfo{volume}{26}},
  \bibinfo{pages}{233201} (\bibinfo{year}{2014}).

\bibitem[{\citenamefont{Cazorla and Boronat}(2017)}]{ca17}
\bibinfo{author}{\bibfnamefont{C.}~\bibnamefont{Cazorla}} \bibnamefont{and}
  \bibinfo{author}{\bibfnamefont{J.}~\bibnamefont{Boronat}},
  \bibinfo{journal}{Rev. Mod. Phys.} \textbf{\bibinfo{volume}{89}},
  \bibinfo{pages}{035003} (\bibinfo{year}{2017}).

\bibitem[{\citenamefont{Los et~al.}(2005)\citenamefont{Los, Ghiringhelli,
  Meijer, and Fasolino}}]{lo05}
\bibinfo{author}{\bibfnamefont{J.~H.} \bibnamefont{Los}},
  \bibinfo{author}{\bibfnamefont{L.~M.} \bibnamefont{Ghiringhelli}},
  \bibinfo{author}{\bibfnamefont{E.~J.} \bibnamefont{Meijer}},
  \bibnamefont{and} \bibinfo{author}{\bibfnamefont{A.}~\bibnamefont{Fasolino}},
  \bibinfo{journal}{Phys. Rev. B} \textbf{\bibinfo{volume}{72}},
  \bibinfo{pages}{214102} (\bibinfo{year}{2005}).

\bibitem[{\citenamefont{Ghiringhelli et~al.}(2005)\citenamefont{Ghiringhelli,
  Los, Meijer, Fasolino, and Frenkel}}]{gh05b}
\bibinfo{author}{\bibfnamefont{L.~M.} \bibnamefont{Ghiringhelli}},
  \bibinfo{author}{\bibfnamefont{J.~H.} \bibnamefont{Los}},
  \bibinfo{author}{\bibfnamefont{E.~J.} \bibnamefont{Meijer}},
  \bibinfo{author}{\bibfnamefont{A.}~\bibnamefont{Fasolino}}, \bibnamefont{and}
  \bibinfo{author}{\bibfnamefont{D.}~\bibnamefont{Frenkel}},
  \bibinfo{journal}{Phys. Rev. Lett.} \textbf{\bibinfo{volume}{94}},
  \bibinfo{pages}{145701} (\bibinfo{year}{2005}).

\bibitem[{\citenamefont{Ram\'irez and Herrero}(2017)}]{ra17}
\bibinfo{author}{\bibfnamefont{R.}~\bibnamefont{Ram\'irez}} \bibnamefont{and}
  \bibinfo{author}{\bibfnamefont{C.~P.} \bibnamefont{Herrero}},
  \bibinfo{journal}{Phys. Rev. B} \textbf{\bibinfo{volume}{95}},
  \bibinfo{pages}{045423} (\bibinfo{year}{2017}).

\bibitem[{\citenamefont{Zakharchenko et~al.}(2009)\citenamefont{Zakharchenko,
  Katsnelson, and Fasolino}}]{za09}
\bibinfo{author}{\bibfnamefont{K.~V.} \bibnamefont{Zakharchenko}},
  \bibinfo{author}{\bibfnamefont{M.~I.} \bibnamefont{Katsnelson}},
  \bibnamefont{and} \bibinfo{author}{\bibfnamefont{A.}~\bibnamefont{Fasolino}},
  \bibinfo{journal}{Phys. Rev. Lett.} \textbf{\bibinfo{volume}{102}},
  \bibinfo{pages}{046808} (\bibinfo{year}{2009}).

\bibitem[{\citenamefont{Politano et~al.}(2012)\citenamefont{Politano, Marino,
  Campi, Far\'ias, Miranda, and Chiarello}}]{po12}
\bibinfo{author}{\bibfnamefont{A.}~\bibnamefont{Politano}},
  \bibinfo{author}{\bibfnamefont{A.~R.} \bibnamefont{Marino}},
  \bibinfo{author}{\bibfnamefont{D.}~\bibnamefont{Campi}},
  \bibinfo{author}{\bibfnamefont{D.}~\bibnamefont{Far\'ias}},
  \bibinfo{author}{\bibfnamefont{R.}~\bibnamefont{Miranda}}, \bibnamefont{and}
  \bibinfo{author}{\bibfnamefont{G.}~\bibnamefont{Chiarello}},
  \bibinfo{journal}{Carbon} \textbf{\bibinfo{volume}{50}}, \bibinfo{pages}{4903
  } (\bibinfo{year}{2012}).

\bibitem[{\citenamefont{Ram\'irez and Herrero}(2018)}]{ra18b}
\bibinfo{author}{\bibfnamefont{R.}~\bibnamefont{Ram\'irez}} \bibnamefont{and}
  \bibinfo{author}{\bibfnamefont{C.~P.} \bibnamefont{Herrero}},
  \bibinfo{journal}{J. Chem. Phys.} \textbf{\bibinfo{volume}{149}},
  \bibinfo{pages}{041102} (\bibinfo{year}{2018}).

\bibitem[{\citenamefont{Ram\'irez et~al.}(2016)\citenamefont{Ram\'irez,
  Chac\'on, and Herrero}}]{ra16}
\bibinfo{author}{\bibfnamefont{R.}~\bibnamefont{Ram\'irez}},
  \bibinfo{author}{\bibfnamefont{E.}~\bibnamefont{Chac\'on}}, \bibnamefont{and}
  \bibinfo{author}{\bibfnamefont{C.~P.} \bibnamefont{Herrero}},
  \bibinfo{journal}{Phys. Rev. B} \textbf{\bibinfo{volume}{93}},
  \bibinfo{pages}{235419} (\bibinfo{year}{2016}).

\bibitem[{\citenamefont{Lambin}(2014)}]{la14}
\bibinfo{author}{\bibfnamefont{P.}~\bibnamefont{Lambin}},
  \bibinfo{journal}{Appl. Sci.} \textbf{\bibinfo{volume}{4}},
  \bibinfo{pages}{282} (\bibinfo{year}{2014}).

\bibitem[{\citenamefont{Tuckerman and Hughes}(1998)}]{tu98}
\bibinfo{author}{\bibfnamefont{M.~E.} \bibnamefont{Tuckerman}}
  \bibnamefont{and} \bibinfo{author}{\bibfnamefont{A.}~\bibnamefont{Hughes}},
  in \emph{\bibinfo{booktitle}{Classical and Quantum Dynamics in Condensed
  Phase Simulations}}, edited by \bibinfo{editor}{\bibfnamefont{B.~J.}
  \bibnamefont{Berne}},
  \bibinfo{editor}{\bibfnamefont{G.}~\bibnamefont{Ciccotti}}, \bibnamefont{and}
  \bibinfo{editor}{\bibfnamefont{D.~F.} \bibnamefont{Coker}}
  (\bibinfo{publisher}{Word Scientific}, \bibinfo{address}{Singapore},
  \bibinfo{year}{1998}), p. \bibinfo{pages}{311}.

\bibitem[{\citenamefont{Martyna et~al.}(1999)\citenamefont{Martyna, Hughes, and
  Tuckerman}}]{ma99}
\bibinfo{author}{\bibfnamefont{G.~J.} \bibnamefont{Martyna}},
  \bibinfo{author}{\bibfnamefont{A.}~\bibnamefont{Hughes}}, \bibnamefont{and}
  \bibinfo{author}{\bibfnamefont{M.~E.} \bibnamefont{Tuckerman}},
  \bibinfo{journal}{J. Chem. Phys.} \textbf{\bibinfo{volume}{110}},
  \bibinfo{pages}{3275} (\bibinfo{year}{1999}).

\bibitem[{\citenamefont{Tuckerman}(2002)}]{tu02}
\bibinfo{author}{\bibfnamefont{M.~E.} \bibnamefont{Tuckerman}}, in
  \emph{\bibinfo{booktitle}{Quantum Simulations of Complex Many--Body Systems:
  From Theory to Algorithms}}, edited by
  \bibinfo{editor}{\bibfnamefont{J.}~\bibnamefont{Grotendorst}},
  \bibinfo{editor}{\bibfnamefont{D.}~\bibnamefont{Marx}}, \bibnamefont{and}
  \bibinfo{editor}{\bibfnamefont{A.}~\bibnamefont{Muramatsu}}
  (\bibinfo{publisher}{NIC}, \bibinfo{address}{FZ J\"ulich},
  \bibinfo{year}{2002}), p. \bibinfo{pages}{269}.

\bibitem[{\citenamefont{Herman et~al.}(1982)\citenamefont{Herman, Bruskin, and
  Berne}}]{he82}
\bibinfo{author}{\bibfnamefont{M.~F.} \bibnamefont{Herman}},
  \bibinfo{author}{\bibfnamefont{E.~J.} \bibnamefont{Bruskin}},
  \bibnamefont{and} \bibinfo{author}{\bibfnamefont{B.~J.} \bibnamefont{Berne}},
  \bibinfo{journal}{J. Chem. Phys.} \textbf{\bibinfo{volume}{76}},
  \bibinfo{pages}{5150} (\bibinfo{year}{1982}).

\bibitem[{\citenamefont{Herrero et~al.}(2006)\citenamefont{Herrero,
  Ram\'{\i}rez, and Hern\'andez}}]{he06}
\bibinfo{author}{\bibfnamefont{C.~P.} \bibnamefont{Herrero}},
  \bibinfo{author}{\bibfnamefont{R.}~\bibnamefont{Ram\'{\i}rez}},
  \bibnamefont{and} \bibinfo{author}{\bibfnamefont{E.~R.}
  \bibnamefont{Hern\'andez}}, \bibinfo{journal}{Phys. Rev. B}
  \textbf{\bibinfo{volume}{73}}, \bibinfo{pages}{245211}
  (\bibinfo{year}{2006}).

\bibitem[{\citenamefont{Herrero and Ram\'irez}(2011)}]{he11}
\bibinfo{author}{\bibfnamefont{C.~P.} \bibnamefont{Herrero}} \bibnamefont{and}
  \bibinfo{author}{\bibfnamefont{R.}~\bibnamefont{Ram\'irez}},
  \bibinfo{journal}{J. Chem. Phys.} \textbf{\bibinfo{volume}{134}},
  \bibinfo{pages}{094510} (\bibinfo{year}{2011}).

\bibitem[{\citenamefont{Ram\'irez et~al.}(2012)\citenamefont{Ram\'irez,
  Neuerburg, Fern\'andez-Serra, and Herrero}}]{ra12}
\bibinfo{author}{\bibfnamefont{R.}~\bibnamefont{Ram\'irez}},
  \bibinfo{author}{\bibfnamefont{N.}~\bibnamefont{Neuerburg}},
  \bibinfo{author}{\bibfnamefont{M.~V.} \bibnamefont{Fern\'andez-Serra}},
  \bibnamefont{and} \bibinfo{author}{\bibfnamefont{C.~P.}
  \bibnamefont{Herrero}}, \bibinfo{journal}{J. Chem. Phys.}
  \textbf{\bibinfo{volume}{137}}, \bibinfo{pages}{044502}
  (\bibinfo{year}{2012}).

\bibitem[{\citenamefont{Waheed and Edholm}(2009)}]{wa09}
\bibinfo{author}{\bibfnamefont{Q.}~\bibnamefont{Waheed}} \bibnamefont{and}
  \bibinfo{author}{\bibfnamefont{O.}~\bibnamefont{Edholm}},
  \bibinfo{journal}{Biophys. J.} \textbf{\bibinfo{volume}{97}},
  \bibinfo{pages}{2754} (\bibinfo{year}{2009}).

\bibitem[{\citenamefont{Pozzo et~al.}(2011)\citenamefont{Pozzo, Alf\`e,
  Lacovig, Hofmann, Lizzit, and Baraldi}}]{po11b}
\bibinfo{author}{\bibfnamefont{M.}~\bibnamefont{Pozzo}},
  \bibinfo{author}{\bibfnamefont{D.}~\bibnamefont{Alf\`e}},
  \bibinfo{author}{\bibfnamefont{P.}~\bibnamefont{Lacovig}},
  \bibinfo{author}{\bibfnamefont{P.}~\bibnamefont{Hofmann}},
  \bibinfo{author}{\bibfnamefont{S.}~\bibnamefont{Lizzit}}, \bibnamefont{and}
  \bibinfo{author}{\bibfnamefont{A.}~\bibnamefont{Baraldi}},
  \bibinfo{journal}{Phys. Rev. Lett.} \textbf{\bibinfo{volume}{106}},
  \bibinfo{pages}{135501} (\bibinfo{year}{2011}).

\bibitem[{\citenamefont{Nicholl et~al.}(2017)\citenamefont{Nicholl, Lavrik,
  Vlassiouk, Srijanto, and Bolotin}}]{ni17}
\bibinfo{author}{\bibfnamefont{R.~J.~T.} \bibnamefont{Nicholl}},
  \bibinfo{author}{\bibfnamefont{N.~V.} \bibnamefont{Lavrik}},
  \bibinfo{author}{\bibfnamefont{I.}~\bibnamefont{Vlassiouk}},
  \bibinfo{author}{\bibfnamefont{B.~R.} \bibnamefont{Srijanto}},
  \bibnamefont{and} \bibinfo{author}{\bibfnamefont{K.~I.}
  \bibnamefont{Bolotin}}, \bibinfo{journal}{Phys. Rev. Lett.}
  \textbf{\bibinfo{volume}{118}}, \bibinfo{pages}{266101}
  (\bibinfo{year}{2017}).

\bibitem[{\citenamefont{Fournier and Barbetta}(2008)}]{fo08}
\bibinfo{author}{\bibfnamefont{J.-B.} \bibnamefont{Fournier}} \bibnamefont{and}
  \bibinfo{author}{\bibfnamefont{C.}~\bibnamefont{Barbetta}},
  \bibinfo{journal}{Phys. Rev. Lett.} \textbf{\bibinfo{volume}{100}},
  \bibinfo{pages}{078103} (\bibinfo{year}{2008}).

\bibitem[{\citenamefont{Nicholl et~al.}(2015)\citenamefont{Nicholl, Conley,
  Lavrik, Vlassiouk, Puzyrev, Sreenivas, Pantelides, and Bolotin}}]{ni15}
\bibinfo{author}{\bibfnamefont{R.~J.~T.} \bibnamefont{Nicholl}},
  \bibinfo{author}{\bibfnamefont{H.~J.} \bibnamefont{Conley}},
  \bibinfo{author}{\bibfnamefont{N.~V.} \bibnamefont{Lavrik}},
  \bibinfo{author}{\bibfnamefont{I.}~\bibnamefont{Vlassiouk}},
  \bibinfo{author}{\bibfnamefont{Y.~S.} \bibnamefont{Puzyrev}},
  \bibinfo{author}{\bibfnamefont{V.~P.} \bibnamefont{Sreenivas}},
  \bibinfo{author}{\bibfnamefont{S.~T.} \bibnamefont{Pantelides}},
  \bibnamefont{and} \bibinfo{author}{\bibfnamefont{K.~I.}
  \bibnamefont{Bolotin}}, \bibinfo{journal}{Nature Commun.}
  \textbf{\bibinfo{volume}{6}}, \bibinfo{pages}{8789} (\bibinfo{year}{2015}).

\bibitem[{\citenamefont{Helfrich and Servuss}(1984)}]{he84}
\bibinfo{author}{\bibfnamefont{W.}~\bibnamefont{Helfrich}} \bibnamefont{and}
  \bibinfo{author}{\bibfnamefont{R.~M.} \bibnamefont{Servuss}},
  \bibinfo{journal}{Nuovo Cimento D} \textbf{\bibinfo{volume}{3}},
  \bibinfo{pages}{137} (\bibinfo{year}{1984}).

\bibitem[{\citenamefont{Herrero and Ram\'irez}(2018{\natexlab{b}})}]{he18b}
\bibinfo{author}{\bibfnamefont{C.~P.} \bibnamefont{Herrero}} \bibnamefont{and}
  \bibinfo{author}{\bibfnamefont{R.}~\bibnamefont{Ram\'irez}},
  \bibinfo{journal}{Phys. Rev. B} \textbf{\bibinfo{volume}{97}},
  \bibinfo{pages}{195433} (\bibinfo{year}{2018}{\natexlab{b}}).

\bibitem[{\citenamefont{Gillan}(1990)}]{gi90}
\bibinfo{author}{\bibfnamefont{M.~J.} \bibnamefont{Gillan}}, in
  \emph{\bibinfo{booktitle}{Computer Modelling of Fluids, Polymers and
  Solids}}, edited by \bibinfo{editor}{\bibfnamefont{C.~R.~A.}
  \bibnamefont{Catlow}}, \bibinfo{editor}{\bibfnamefont{S.~C.}
  \bibnamefont{Parker}}, \bibnamefont{and}
  \bibinfo{editor}{\bibfnamefont{M.~P.} \bibnamefont{Allen}}
  (\bibinfo{publisher}{Kluwer}, \bibinfo{address}{Dordrecht},
  \bibinfo{year}{1990}), p. \bibinfo{pages}{155}.

\bibitem[{\citenamefont{Karssemeijer and Fasolino}(2011)}]{ka11}
\bibinfo{author}{\bibfnamefont{L.~J.} \bibnamefont{Karssemeijer}}
  \bibnamefont{and} \bibinfo{author}{\bibfnamefont{A.}~\bibnamefont{Fasolino}},
  \bibinfo{journal}{Surf. Sci.} \textbf{\bibinfo{volume}{605}},
  \bibinfo{pages}{1611} (\bibinfo{year}{2011}).

\bibitem[{\citenamefont{Yan et~al.}({2008})\citenamefont{Yan, Ruan, and
  Chou}}]{ya08}
\bibinfo{author}{\bibfnamefont{J.-A.} \bibnamefont{Yan}},
  \bibinfo{author}{\bibfnamefont{W.~Y.} \bibnamefont{Ruan}}, \bibnamefont{and}
  \bibinfo{author}{\bibfnamefont{M.~Y.} \bibnamefont{Chou}},
  \bibinfo{journal}{Phys. Rev. B} \textbf{\bibinfo{volume}{{77}}},
  \bibinfo{pages}{{125401}} (\bibinfo{year}{{2008}}).

\bibitem[{\citenamefont{Singh and Hennig}({2013})}]{si13b}
\bibinfo{author}{\bibfnamefont{A.~K.} \bibnamefont{Singh}} \bibnamefont{and}
  \bibinfo{author}{\bibfnamefont{R.~G.} \bibnamefont{Hennig}},
  \bibinfo{journal}{Phys. Rev. B} \textbf{\bibinfo{volume}{{87}}},
  \bibinfo{pages}{{094112}} (\bibinfo{year}{{2013}}).

\bibitem[{\citenamefont{Koukaras et~al.}({2015})\citenamefont{Koukaras,
  Kalosakas, Galiotis, and Papagelis}}]{ko15b}
\bibinfo{author}{\bibfnamefont{E.~N.} \bibnamefont{Koukaras}},
  \bibinfo{author}{\bibfnamefont{G.}~\bibnamefont{Kalosakas}},
  \bibinfo{author}{\bibfnamefont{C.}~\bibnamefont{Galiotis}}, \bibnamefont{and}
  \bibinfo{author}{\bibfnamefont{K.}~\bibnamefont{Papagelis}},
  \bibinfo{journal}{Sci. Rep.} \textbf{\bibinfo{volume}{{5}}},
  \bibinfo{pages}{{12923}} (\bibinfo{year}{{2015}}).

\bibitem[{\citenamefont{Zakharchenko et~al.}(2010)\citenamefont{Zakharchenko,
  Los, Katsnelson, and Fasolino}}]{za10b}
\bibinfo{author}{\bibfnamefont{K.~V.} \bibnamefont{Zakharchenko}},
  \bibinfo{author}{\bibfnamefont{J.~H.} \bibnamefont{Los}},
  \bibinfo{author}{\bibfnamefont{M.~I.} \bibnamefont{Katsnelson}},
  \bibnamefont{and} \bibinfo{author}{\bibfnamefont{A.}~\bibnamefont{Fasolino}},
  \bibinfo{journal}{Phys. Rev. B} \textbf{\bibinfo{volume}{81}},
  \bibinfo{pages}{235439} (\bibinfo{year}{2010}).

\bibitem[{\citenamefont{Yamanaka et~al.}(1994)\citenamefont{Yamanaka, Morimoto,
  and Kanda}}]{ya94}
\bibinfo{author}{\bibfnamefont{T.}~\bibnamefont{Yamanaka}},
  \bibinfo{author}{\bibfnamefont{S.}~\bibnamefont{Morimoto}}, \bibnamefont{and}
  \bibinfo{author}{\bibfnamefont{H.}~\bibnamefont{Kanda}},
  \bibinfo{journal}{Phys. Rev. B} \textbf{\bibinfo{volume}{49}},
  \bibinfo{pages}{9341} (\bibinfo{year}{1994}).

\bibitem[{\citenamefont{Ramdas et~al.}(1993)\citenamefont{Ramdas, Rodriguez,
  Grimsditch, Anthony, and Banholzer}}]{ra93b}
\bibinfo{author}{\bibfnamefont{A.~K.} \bibnamefont{Ramdas}},
  \bibinfo{author}{\bibfnamefont{S.}~\bibnamefont{Rodriguez}},
  \bibinfo{author}{\bibfnamefont{M.}~\bibnamefont{Grimsditch}},
  \bibinfo{author}{\bibfnamefont{T.~R.} \bibnamefont{Anthony}},
  \bibnamefont{and} \bibinfo{author}{\bibfnamefont{W.~F.}
  \bibnamefont{Banholzer}}, \bibinfo{journal}{Phys. Rev. Lett.}
  \textbf{\bibinfo{volume}{71}}, \bibinfo{pages}{189} (\bibinfo{year}{1993}).

\bibitem[{\citenamefont{Kazimorov et~al.}(1998)\citenamefont{Kazimorov,
  Zegenhagen, and Cardona}}]{ka98}
\bibinfo{author}{\bibfnamefont{A.}~\bibnamefont{Kazimorov}},
  \bibinfo{author}{\bibfnamefont{J.}~\bibnamefont{Zegenhagen}},
  \bibnamefont{and} \bibinfo{author}{\bibfnamefont{M.}~\bibnamefont{Cardona}},
  \bibinfo{journal}{Science} \textbf{\bibinfo{volume}{282}},
  \bibinfo{pages}{930} (\bibinfo{year}{1998}).

\bibitem[{\citenamefont{Kittel}(1966)}]{ki66}
\bibinfo{author}{\bibfnamefont{C.}~\bibnamefont{Kittel}},
  \emph{\bibinfo{title}{Introduction to Solid State Physics}}
  (\bibinfo{publisher}{Wiley}, \bibinfo{address}{New York},
  \bibinfo{year}{1966}).

\bibitem[{\citenamefont{Ashcroft and Mermin}(1976)}]{as76}
\bibinfo{author}{\bibfnamefont{N.~W.} \bibnamefont{Ashcroft}} \bibnamefont{and}
  \bibinfo{author}{\bibfnamefont{N.~D.} \bibnamefont{Mermin}},
  \emph{\bibinfo{title}{Solid State Physics}} (\bibinfo{publisher}{Saunders
  College}, \bibinfo{address}{Philadelphia}, \bibinfo{year}{1976}).

\bibitem[{\citenamefont{Michel et~al.}(2015)\citenamefont{Michel, Costamagna,
  and Peeters}}]{mi15b}
\bibinfo{author}{\bibfnamefont{K.~H.} \bibnamefont{Michel}},
  \bibinfo{author}{\bibfnamefont{S.}~\bibnamefont{Costamagna}},
  \bibnamefont{and} \bibinfo{author}{\bibfnamefont{F.~M.}
  \bibnamefont{Peeters}}, \bibinfo{journal}{Phys. Status Solidi B}
  \textbf{\bibinfo{volume}{252}}, \bibinfo{pages}{2433} (\bibinfo{year}{2015}).

\bibitem[{\citenamefont{Mounet and Marzari}(2005)}]{mo05}
\bibinfo{author}{\bibfnamefont{N.}~\bibnamefont{Mounet}} \bibnamefont{and}
  \bibinfo{author}{\bibfnamefont{N.}~\bibnamefont{Marzari}},
  \bibinfo{journal}{Phys. Rev. B} \textbf{\bibinfo{volume}{71}},
  \bibinfo{pages}{205214} (\bibinfo{year}{2005}).

\bibitem[{\citenamefont{Mostaani et~al.}({2015})\citenamefont{Mostaani,
  Drummond, and Fal'ko}}]{mo15b}
\bibinfo{author}{\bibfnamefont{E.}~\bibnamefont{Mostaani}},
  \bibinfo{author}{\bibfnamefont{N.~D.} \bibnamefont{Drummond}},
  \bibnamefont{and} \bibinfo{author}{\bibfnamefont{V.~I.}
  \bibnamefont{Fal'ko}}, \bibinfo{journal}{Phys. Rev. Lett.}
  \textbf{\bibinfo{volume}{{115}}}, \bibinfo{pages}{{115501}}
  (\bibinfo{year}{{2015}}).

\bibitem[{\citenamefont{Herrero and Ram\'irez}(2000)}]{he00c}
\bibinfo{author}{\bibfnamefont{C.~P.} \bibnamefont{Herrero}} \bibnamefont{and}
  \bibinfo{author}{\bibfnamefont{R.}~\bibnamefont{Ram\'irez}},
  \bibinfo{journal}{Phys. Rev. B} \textbf{\bibinfo{volume}{63}},
  \bibinfo{pages}{024103} (\bibinfo{year}{2000}).

\bibitem[{\citenamefont{Baskin and Meyer}({1955})}]{ba55}
\bibinfo{author}{\bibfnamefont{Y.}~\bibnamefont{Baskin}} \bibnamefont{and}
  \bibinfo{author}{\bibfnamefont{L.}~\bibnamefont{Meyer}},
  \bibinfo{journal}{{Phys. Rev.}} \textbf{\bibinfo{volume}{{100}}},
  \bibinfo{pages}{{544}} (\bibinfo{year}{{1955}}).

\bibitem[{\citenamefont{Callen}(1960)}]{ca60}
\bibinfo{author}{\bibfnamefont{H.~B.} \bibnamefont{Callen}},
  \emph{\bibinfo{title}{Thermodynamics}} (\bibinfo{publisher}{John Wiley},
  \bibinfo{address}{New York}, \bibinfo{year}{1960}).

\bibitem[{\citenamefont{Landau and Lifshitz}(1980)}]{la80}
\bibinfo{author}{\bibfnamefont{L.~D.} \bibnamefont{Landau}} \bibnamefont{and}
  \bibinfo{author}{\bibfnamefont{E.~M.} \bibnamefont{Lifshitz}},
  \emph{\bibinfo{title}{Statistical Physics}} (\bibinfo{publisher}{Pergamon},
  \bibinfo{address}{Oxford}, \bibinfo{year}{1980}), \bibinfo{edition}{3rd} ed.

\bibitem[{\citenamefont{Herrero}(2008)}]{he08}
\bibinfo{author}{\bibfnamefont{C.~P.} \bibnamefont{Herrero}},
  \bibinfo{journal}{J. Phys.: Condens. Matter} \textbf{\bibinfo{volume}{20}},
  \bibinfo{pages}{295230} (\bibinfo{year}{2008}).

\bibitem[{\citenamefont{Blakslee et~al.}({1970})\citenamefont{Blakslee,
  Proctor, Seldin, Spence, and Weng}}]{bl70}
\bibinfo{author}{\bibfnamefont{O.~L.} \bibnamefont{Blakslee}},
  \bibinfo{author}{\bibfnamefont{D.~G.} \bibnamefont{Proctor}},
  \bibinfo{author}{\bibfnamefont{E.~J.} \bibnamefont{Seldin}},
  \bibinfo{author}{\bibfnamefont{G.~B.} \bibnamefont{Spence}},
  \bibnamefont{and} \bibinfo{author}{\bibfnamefont{T.}~\bibnamefont{Weng}},
  \bibinfo{journal}{{J. Appl. Phys.}} \textbf{\bibinfo{volume}{{41}}},
  \bibinfo{pages}{{3373}} (\bibinfo{year}{{1970}}).

\bibitem[{\citenamefont{Nicklow et~al.}(1972)\citenamefont{Nicklow,
  Wakabayashi, and Smith}}]{ni72}
\bibinfo{author}{\bibfnamefont{R.}~\bibnamefont{Nicklow}},
  \bibinfo{author}{\bibfnamefont{N.}~\bibnamefont{Wakabayashi}},
  \bibnamefont{and} \bibinfo{author}{\bibfnamefont{H.~G.} \bibnamefont{Smith}},
  \bibinfo{journal}{Phys. Rev. B} \textbf{\bibinfo{volume}{5}},
  \bibinfo{pages}{4951} (\bibinfo{year}{1972}).

\bibitem[{\citenamefont{Lin et~al.}(2018)\citenamefont{Lin, Wu, Liu, and
  Tan}}]{li18}
\bibinfo{author}{\bibfnamefont{M.-L.} \bibnamefont{Lin}},
  \bibinfo{author}{\bibfnamefont{J.-B.} \bibnamefont{Wu}},
  \bibinfo{author}{\bibfnamefont{X.-L.} \bibnamefont{Liu}}, \bibnamefont{and}
  \bibinfo{author}{\bibfnamefont{P.-H.} \bibnamefont{Tan}},
  \bibinfo{journal}{J. Raman Spectr.} \textbf{\bibinfo{volume}{49}},
  \bibinfo{pages}{19} (\bibinfo{year}{2018}).

\bibitem[{\citenamefont{Komatsu}({1964})}]{ko64}
\bibinfo{author}{\bibfnamefont{K.}~\bibnamefont{Komatsu}}, \bibinfo{journal}{J.
  Phys. Chem. Solids} \textbf{\bibinfo{volume}{{25}}}, \bibinfo{pages}{707}
  (\bibinfo{year}{{1964}}).

\bibitem[{\citenamefont{Herrero and Ram\'irez}(2017)}]{he17}
\bibinfo{author}{\bibfnamefont{C.~P.} \bibnamefont{Herrero}} \bibnamefont{and}
  \bibinfo{author}{\bibfnamefont{R.}~\bibnamefont{Ram\'irez}},
  \bibinfo{journal}{Phys. Chem. Chem. Phys.} \textbf{\bibinfo{volume}{19}},
  \bibinfo{pages}{31898} (\bibinfo{year}{2017}).

\end{thebibliography}
\end{document}